# ILETIA: An AI-enhanced method for individualized trigger-oocyte pickup interval estimation of progestin-primed ovarian stimulation protocol

Binjian Wu[1,2,#], Qian Li[2,#], Zhe Kuang[1,#], Hongyuan Gao[1], Xinyi Liu[2], Haiyan Guo[1], Qiuju Chen[1], Xinyi Liu[1], Yangruizhe Jiang[3], Yuqi Zhang[1], Jinyin Zha[2], Mingyu Li[2], Qiuhan Ren[2], Sishuo Feng[2], Haicang Zhang[2], Xuefeng Lu[1,*], Jian Zhang[2,*]

[1] Department of Assisted Reproduction, Shanghai Ninth People's Hospital, Shanghai Jiao Tong University School of Medicine (SJTU-SM), Shanghai, 200011, China
[2] Medicinal Bioinformatics Center, Shanghai Jiao Tong University School of Medicine (SJTU-SM), Shanghai, 200025, China
[3] School of Computing Science, University of Glasgow, Glasgow, Glasgow, United Kingdom

[#] The authors equally contribute to this work
[*] Corresponding author
Dr. Xuefeng Lu
**Email**: xuefenglu163@163.com
Dr. Jian Zhang
**Email**: jian.zhang@sjtu.edu.cn

# Abstract

*in vitro* fertilization-embryo transfer (IVF-ET) stands as one of the most prevalent treatments for infertility. During an IVF-ET cycle, the time interval between trigger shot and oocyte pickup (OPU) is a pivotal period for follicular maturation, which determines mature oocytes yields and impacts the success of subsequent procedures. However, accurately predicting this interval is severely hindered by the variability of clinicians' experience that often leads to suboptimal oocyte retrieval rate. To address this challenge, we propose ILETIA, the first machine learning-based method that could predict the optimal trigger-OPU interval for patients receiving progestin-primed ovarian stimulation (PPOS) protocol. Specifically, ILETIA leverages a Transformer to learn representations from clinical tabular data, and then employs gradient-boosted trees for interval prediction. For model training and evaluating, we compiled a dataset PPOS-DS of nearly ten thousand patients receiving PPOS protocol, the largest such dataset to our knowledge. Experimental results demonstrate that our method achieves strong performance (AUROC = 0.889), outperforming both clinicians and other widely used computational models. Moreover, ILETIA also supports premature ovulation risk prediction in a specific OPU time (AUROC = 0.838). Collectively, by enabling more precise and individualized decisions, ILETIA has the potential to improve clinical outcomes and lay the foundation for future IVF-ET research.

# Introduction

Infertility, a disease of the reproductive system, defined as the failure to achieve a clinical pregnancy after 12 months or more of regular unprotected sexual intercourse [1], can cause significant damage to people's mental and psychosocial well-being. With the development of society, human fertility has gradually declined, with global infertility rates reaching 17.5% estimated by the World Health Organization [2]. The efficiency of *in vitro* fertilization-embryo transfer (IVF-ET) urgently requires improvement. An IVF-ET cycle generally involves five key steps [3]: 1) controlled ovarian hyperstimulation (COH), 2) oocyte retrieval or oocyte pickup (OPU), 3) fertilization, 4) embryo culture and 5) embryo transfer (Fig. 1a). COH is a complicated process where the patient's luteinizing hormone (LH) surge is suppressed by specific protocols temporarily, as the ovaries are stimulated through the injection or oral of specific drugs to produce multiple follicles. Once the follicles reach an appropriate size, a trigger shot is administered to induce the final maturation of the oocytes.

The time interval between ovulation triggering and OPU plays a crucial role in IVF-ET success, as this stage involves several key processes including the start of luteinization, the expansion of the cumulus cells, the resumption of the reduction division of the oocyte and the maturation of oocyte [4]. A time interval that is too short may lead to a low oocyte retrieval rate or empty follicle syndrome, while a time interval that is too long may result in premature ovulation. Premature ovulation is a rare phenomenon with an estimated incidence of 0.34% [5], referring to a dramatic advancement of the actual OPU time compared to the predetermined one. An inappropriate trigger-OPU interval decreases the live birth rate in an IVF-ET cycle [6]. Traditionally, the trigger-OPU interval has relied mainly on the clinical judgment of doctors. Nevertheless, due to individual variability among patients and the influence of multiple clinical factors, a suboptimal oocyte retrieval rate often occurs. Therefore, there is a pressing need to develop an individualized method to predict the optimal trigger-OPU interval, maximizing the oocyte retrieval rate and enhancing the overall efficiency of IVF-ET treatment.

Recent studies have offered valuable insights into estimating the trigger-OPU interval. Several retrospective studies have provided an approximate range or possible optimization strategies of the ideal intervals for commonly used COH protocols [6,7,8,9,10,11,12,13]. Other research has explored the relationship between oocyte mature and various clinical factors, e.g., follicular fluid lipid metabolites [14], follicle size [15,16,17], estradiol (E2) [18], ovarian response [19,20,21], etc. While these works have enriched clinicians' understanding of the trigger-OPU interval to some extent, it is usually ineffective to rely solely on clinical expertise, and it still remains challenging to tailor a patient-specific and accurate trigger-OPU interval. Apart from this, existing research tends to emphasis the isolated effect of individual clinical variables, neglecting that the optimal interval is determined by a comprehensive effect of multiple factors. With regard to the research on premature ovulation, an individualized method for premature ovulation prediction is still lacking though some improved treatments have been proposed [22,23,24,25]. Therefore, current methods fall short in meeting the precise medical needs of patients undergoing IVF-ET.

With the rapid advancement of medical artificial intelligence [26], increasing importance has been attached to applying artificial intelligence (AI) to IVF-ET, yielding encouraging results. Convolutional neural networks (CNN), as one of the most powerful image feature extraction technologies, have shown great potential in embryo selection [27,28,29,30]. A series of traditional machine learning (ML) classifiers have been employed to predict live birth through clinical variables [31,32]. In addition, several AI-driven methods have been applied to customize the initial dose of FSH in order to reduce the risk of both low ovarian response and OHSS [33,34,35]. Likewise, ML methods have demonstrated strong capabilities in predicting the optimal trigger day of COH [36] [37], and could enhance clinicians' confidence with interpretable technology [37]. Beyond these applications, medical AI has also shown significant potential in terms of predicting pregnancy [38,39,40], oocytes numbers retrieved [41], oocyte maturation rate [42], high-quality sperm [43] and early pregnancy loss [44]. Presently, it still remains unexplored to leverage AI to investigate the optimal trigger-OPU interval.

The progestin-primed ovarian stimulation (PPOS) protocol is a novel COH protocol that first proposed by Kuang et al. in 2015 using progesin, such as medroxyprogesterone acetate (MPA), as a viable substitute for traditional GnRH analogs [45]. In this protocol, progesin exerts a negative feedback effect on the hypothalamus and pituitary gland, inhibiting the secretion of FSH and LH. PPOS protocol offers a variety of advantages. It demonstrates a more stable suppression of a premature luteinizing. The oral administration of progesin and low cost enhances the convenience for clinical practice. Moreover, it has been proven to decrease the risk of ovarian hyperstimulation syndrome (OHSS) [45,46,47]. For these reasons, PPOS protocol has gained considerable prevalence nowadays.

Herein we compile a dataset PPOS-DS involving 8,730 patients receiving PPOS protocol, the largest such dataset to our knowledge, and develop an AI-enhanced method ILETIA for individualized trigger-OPU interval estimation. ILETIA consists of 2 parallel functional modules: the optimal trigger-OPU interval predictors and the premature ovulation risk predictor. The former comprises a 3-class classifier and a regressor outputting an interval rank and an interval value respectively. Both the models ensemble a Transformer [48] and a gradient-boosted trees (GBT): the classifier FTXGB is constructed by FT Transformer architecture [49] integrated with XGBoost [50], while the regressor FTLGB by FT Transformer architecture integrated with LightGBM [51] (Fig. 1b). This construction aims to harness the powerful feature extraction capability of Transformer to capture the latent interactions of clinical variables, and the accurate prediction capability of GBT to avoid overfitting. The latter is designed to predict the premature ovulation probability in an artificially estimated OPU time if clinicians tend to predict the interval based on their own experience, which is built on XGBoost and utilizes CTGAN for oversampling [52]. Model interpretability is implemented by SHAP [53] so as to illustrate how clinical variables influence the final output. To the best of our knowledge, this is the first research of an AI-based method for the optimal trigger-OPU interval estimation.

# Results

## Study design

ILETIA ( https://mdl.shsmu.edu.cn/ILETIA/ ) estimates the trigger-OPU interval of PPOS protocol according to clinical variables, which consists of the optimal trigger-OPU interval predictors (hereinafter referred to as the interval predictors) and the premature ovulation risk predictor (hereinafter referred to the risk predictor). This tool is developed based on the dataset PPOS-DS that comprises clinical information of 8,730 anonymized patients who received PPOS protocol between December 2013 and July 2024 at our center (Supplementary Fig. 1a).

To construct the interval predictors, we created a dataset of 8,657 cases (Dataset A) based on PPOS-DS for model training and evaluation, and relabeled them by linearly splitting the cases into 3 ranks according to their original continuous label values, i.e., Early, Mid and Late (Supplementary Fig. 1b). Subsequently, we designed a 3-class classifier FTXGB for predicting an interval rank, which was trained on Dataset A's training set (Methods). We also built a regressor FTLGB trained with the original labels for predicting an interval value (Methods). The classifier and the regressor work together as a unified module in parallel.

The risk predictor was treated as a rare event binary classification where patients with premature ovulation in the OPU time were considered positive cases, and those without premature ovulation were considered negative cases. To construct this module, we created another dataset of 800 cases as Dataset B (Supplementary Fig. 1c), on which a XGBoost binary classifier was trained to assess the premature ovulation.

## 3-class classification for the optimal trigger-OPU interval

Both 10-fold cross-validation and full training on Dataset A's training set were conducted for 3-class classification. The result of the cross-validation was calculated as the average of all folds. The performance was compared with that of 5 traditional ML models and 6 deep tabular models (Table 1). Among these models, FTXGB achieved the best performance on 5 out of the 7 metrics in the cross-validation and 6 out of the 7 metrics on Dataset A's test set. We also found that the other models performed worse in the cross-validation than on the Dataset A test set as most metrics declined. In contrast FTXGB still maintained strong performance (accuracy = 0.821, f1 score = 0.782, AUROC = 0.889).

Furthermore, the Receiver Operating Characteristic (ROC) for each rank in the cross-validation of FTXGB indicated that our model exhibits relatively strong stability against data fluctuations (Fig. 2a-c). From the ROC for each rank across all 12 models on Dataset A's test set (Fig. 2d-f), FTXGB demonstrated strong performance across all 3 ranks (AUROC for Early rank: 0.906, AUROC for Mid rank: 0.812, AUROC for Late rank: 0.948), suggesting minimal prediction bias toward any specific rank.

Additionally, we noticed that the performances of GBTs were parallel to those of the other deep tabular models [54], which is consistent with existing studies [54,55], indicating that FTXGB could generate a complementary effect and outperform solely a trained GBT or deep tabular model. We also conducted label randomization experiment with AUROC of 0.501 in the cross-validation and of 0.449 on Dataset A's test set. In summary, FTXGB achieved promising prediction performances across patients of all interval ranks.

## Regression for the optimal trigger-OPU interval

The regressor FTLGB outputs a trigger-OPU interval value to enable quantitative prediction. We conducted 10-fold cross-validation and full training on Dataset A's training set as in the previous classification task, and the results were compared with those of Elastic Net (EN), MLP, Support Vector Machine (SVM), CatBoost [56], XGBoost and LightGBM. FTLGB outperformed all the other models both in the cross-validation (RMSE = 24.85, MAE = 19.56, $r$ = 0.793) and on the test set (RMSE = 20.01, MAE = 16.70, $r$ = 0.747) across the 3 metrics, shown in Table 2 and Fig. 2g. This suggested that Transformer architecture contributed to capturing more latent relationships between clinical variables and LightGBM could reduce overfitting in deep neural networks.

We also observed that MLP performed considerably poorly in the cross-validation, suggesting that deep tabular models tended to overfit on Dataset A, thus we did not evaluate MLP on Dataset A's test set, nor did we try other deep tabular models. Generally, FTLGB still exhibits promise for further applications despite the moderate performance, as its MAE (16.70) is clinically tolerable in PPOS protocol.

## Correlation between optimal trigger-OPU interval rate and high-quality embryo rate, mature oocyte and fertilized oocyte rate

High-quality embryo rate, mature oocyte rate and fertilized oocyte rate are also key factors for IVF-ET success. We investigated whether the optimal trigger-OPU interval is associated with these factors. Spearman correlation coefficients revealed that the optimal trigger-OPU interval was not significantly correlated with high-quality embryo rate (r = -0.07, p <$10^{-11}$), mature oocyte rate (r = -0.05, p < $10^{-6}$), or fertilized oocyte rate (r = -0.06, p < $10^{-7}$). These results ensured our prediction would only affect the oocyte retrieval rate.

## Premature ovulation risk prediction

Since premature ovulation is a rare event during the IVF-ET treatment (we collected a total of 95 cases between July 2015 and July 2023), we utilized CTGAN for oversampling on a subset (randomly selected 55 cases) of all positive cases and synthesized 305 new positive cases of high quality (average KSComplement = 0.800, more details shown in Supplementary Table. 3), and then constructed a balanced Dataset B with a training set containing 720 cases and a test set containing 80 cases (Methods, Supplementary Fig. 1c). Subsequently we developed an XGBoost binary classifier as the premature ovulation risk predictor and achieved high performance on Dataset B's test set with an AUROC of 0.838 and an accuracy of 0.825 (Fig. 2h, Supplementary Table 9).

## Effect of clinical variables on the trigger-OPU interval.

We employed SHAP for model interpretability so as to illustrate feature importance and explore how it influenced the model's decision. The interpretation of FTXGB (Fig. 2j-l) suggested that for top-15 features, their impact on the classification prediction was consistent with existing research and clinical experience. Notably, E2 ratio emerged as the most predictive feature for determining the optimal trigger-OPU interval, with a lower feature value indicating a shorter interval and a higher value indicating a longer one. This was in accordance with previous studies [18,42], which demonstrated that the decrease in preovulatory serum E2 acted as a useful marker indicating an earlier ovulation in natural or unstimulated IVF-ET cycles. hCG dose was also of great consequence to the outcome with a higher level resulting in a shorter interval, as it mimicked the endogenous LH surge to induce final oocyte maturation [57]. Several features contributed moderately to the prediction. The number of follicles of different sizes followed a similar trend that a higher number of follicles generally corresponds to a longer interval and vice versa for a lower number, though the degree of their impacts varied by size. Consistent with this finding, existing research has identified follicles ranging from 12 to 19 mm on trigger day as a key factor for the number of oocytes retrieved [17]. Moreover, AFC was correlated with the optimal interval, especially for Late rank, where a higher number led to a longer interval. This has been confirmed by previous research indicating that AFC could work as a marker for ovarian response [19] and a patient with AFC less than 5 may have diminished ovarian reserve during an IVF-ET cycle [58]. Additionally, patient's age could also influence the output to some extent [59].

As for the interpretation of FTLGB (Supplementary Fig. 4j), the feature importance ranking showed little fluctuations compared to that of FTXGB, and the impact of the same feature on the output remained largely consistent, which indicated that the two models exhibited a similar decision-making mechanism.

We noticed that the interpretation of premature ovulation risk predictor (Supplementary Fig. 4k) did not show a similar feature importance ranking to FTXGB and FTLGB, and most clinical features did not exhibit a clear relationship to the prediction. This could be attributed to the limited samples in Dataset B's test set. Despite this, the top 2 features, E2 ratio and AFC, still demonstrated a relatively significant impact on the output and were consistent with previous studies [18,19,42,58]. In conclusion, the interpretability of our methods could bridge the gap in experience among clinicians and enhance clinical decision-making.

## Comparison of ILETIA with IVF-ET specialists

To compare the prediction capability of ILETIA's interval predictor FTXGB with that of IVF-ET specialists, we recruited 4 attending physicians with over 10 years of experience performing OPU from our center. These specialists were requested to independently predict the optimal trigger-OPU interval rank of 200 patients with no time limit. These patients were randomly selected from Dataset A's test set, and the specialists had no prior involvement in their treatment, encountering them for the first time during the study. The information provided to the specialists included only 31 clinical variables used as input of FTXGB.

The result indicated that our model outperformed IVF-ET specialists for patients in Early or Late rank, while achieving comparable performance for those in Mid rank (Fig. 3a). Specifically, for Early rank, our model achieved the highest AUROC of 0.899, compared to the specialists' best of 0.857. For Mid rank, both our model and the top specialist performed closely, with AUROCs of 0.797 and 0.802, respectively. For Late rank, our model again performed the best with an AUROC of 0.946, exceeding the specialists' best of 0.9.

## Case analysis of the second oocyte retrieval

The second oocyte retrieval refers to the procedure in which, when a patient's oocyte retrieval rate is low in the initial OPU, the clinician may extend the initial trigger-OPU interval appropriately and perform OPU a second time to increase oocyte retrieval rate. For these patients, the optimal trigger-OPU interval is obviously longer than the initial interval. We conducted an analysis of the second oocyte retrieval to examine if the prediction by our method was consistent with or closer to the second one. We collected 33 patients who underwent the second oocyte retrieval of PPOS protocol from December 2017 to March 2023, predicted their optimal interval using FTXGB, and then compared the results with both the initial and second intervals. The predictions of 26 cases (78.8%) demonstrated strong consistency with the second interval (Supplementary Table 4). We also observed that the predictions of 4 cases showed a certain probability both in Mid and Late rank and little probability in Early rank, suggesting that their optimal intervals longer than the median of Mid rank (2,190 min). We also considered their predictions closer to the second interval as their initial intervals were no longer than 2,190 min. Only the results of the classification were presented as a large proportion of the second intervals exceeded 2,280 min, the maximum label of Dataset A's training set.

We also present a specific case to illustrate how ILETIA supports clinical decision-making (Fig. 3b-d). The initial interval was 2,166 min, which resulted in a failed oocyte retrieval rate (0.167). After extending the interval to 2,250 min, the surgery was improved (0.692). ILETIA predicted an interval value of 2,232 min and rank probabilities of 0.0% for Early, 0.5% for Mid, and 99.5% for Late, both in line with the actual second interval (Fig. 3c). The waterfall plot demonstrated that clinical variables related to E2, hCG, BMI, and follicles significantly influenced the prediction, with most of them contributing to increasing the time interval (Fig. 3d). In agreement with these results, we further observed that the patient's feature values were expected to result a long interval. Initially, the patient had a E2 ratio of 1.1, indicating an increased E2 level after the trigger day than that on the trigger day [18]. Then, the total hCG dose she received was relatively low (2000 IU) and below average of Dataset A's training set (2708 IU) [57]. Aside from this, this woman was in class I obesity with a BMI of 33.87 kg/m$^2$ [60]. Lastly, the numbers of follicles over 10 mm (15), 12 mm (15) and 16 mm (12) were obviously greater than those in normal situation [17]. The definition of each feature name is provided in Supplementary Table 1. To conclude, ILETIA could help clinicians make prediction closer to the ideal interval on the initial attempt, avoiding treatment cycle extensions.

## Subgroup analysis

We conducted subgroup analysis of 11 important clinical variables on Dataset A's test set. For each of the variables, the dataset was divided into 2 subgroups, which were subsequently evaluated by FTXGB (Supplementary Table 5). The results demonstrated that macro-average AUROCs of all subgroups were almost more than 0.85 and macro-average AUPRs almost more than 0.75, indicating that our model performed well on these subgroups. Meanwhile, the differences of macro-average AUROCs between the two subgroups of a specific clinical variable were quite low (almost lower than 0.05), indicating good robustness across diverse patient populations. Additionally, we analyzed the subgroups of six variables of clinical interest (Age, Infertility duration, BMI, Duration of stimulation, MPA dose, AFC) in each rank. Our model showed strong robustness with AUROCs almost more than 0.80 and AUPRs almost than 0.75 in each rank (Fig. 4a).

## Research on other COH protocols

To demonstrate the general utility of our method, we constructed 4 datasets of patients receiving four other COH protocols at our center from September 2006 to August 2023. These included: 1,533 cases from the GnRH antagonist protocol, 2,587 cases from the long protocol, 2,103 cases from the short protocol, and 2,746 cases from the mild stimulation protocol. The clinical variables selected for each protocol were similar to those for PPOS but with slight variations in medication-relevant features (details in Supplementary Table 1). For each protocol, regression and classification tasks were conducted by 10-fold cross-validation separately. The relabeling method still linearly categorized the dataset of a specific protocol into 3 ranks (Early, Mid, and Late), but the coverage of each rank differed across protocols (Methods), as it has been reported that the optimal trigger-OPU interval varies significantly among different COH protocols [7,6], and our statistical analysis supported this finding (Supplementary Table 8, Supplementary Fig. 6).

In the regression task, FTLGB outperformed the other models across all metrics and demonstrated moderate or higher performance in 3 out of 4 protocols (Supplementary Table 6). In 3-class classification task, FTXGB demonstrated strong classification performance with macro-average AUROCs above 0.8 across all 4 protocols, as shown in Table 3. We observed that the performance of FTXGB across the four protocols did not exceed that of all the other models as it did in PPOS protocol. FTXGB performed best in the GnRH antagonist protocol and outperformed deep tabular models in the other 3 protocols, where GBTs performed the best. This could be ascribed to insufficient cases of the 4 protocols which lead to overfitting in deep neural networks as our Transformer architecture exhibits a strong dependence on extensive data. Fig. 4b further demonstrated that FTXGB performed well with AUROCs close to 0.9 in Early and Late rank across all 4 protocols, though performance slightly decreased in Mid rank. In general, these results underscored the potential of our method for broader clinical applications.

## Discussion

In this study, we developed an AI-based tool for individualized estimation of oocyte trigger-OPU interval for infertile patients receiving PPOS protocol, namely ILETIA, for the first time in the field of IVF-ET. ILETIA (**I**ndividu**L**ized oocyte retrieval **E**stimator based on gradient-boosted **T**rees **I**ntegrated with **A**ttentions) is adapted from Eileithyia, the Greek goddess of childbirth and midwifery, with the hope of helping infertile families achieve parenthood. Using clinical variables, ILETIA can not only recommend both an interval rank and an interval value for the optimal trigger-OPU interval, but also conduct a premature ovulation assessment.

The time interval from trigger shot to OPU is vital during an IVF-ET cycle as it influences the retrieval rate of mature oocytes [6], thus increasingly drawing research interest of clinicians. Existing research primarily utilizes statistical analyses such as meta-analysis, ANOVA, Chi-squared tests, and Fisher's exact test to provide an approximate range of trigger-OPU intervals for a specific COH protocol [7,6, 8,10,12] or to investigate one specific clinical variable predictive of ovulation[1,14,15,16,18]. These studies have limited clinical applicability as they overlook individual patient variability and the multiplicity of contributing clinical variables. ILETIA is capable of estimating the optimal trigger-OPU interval individually and precisely by concurrently predicting an interval rank and interval value, and is suitable for the majority of infertile women receiving PPOS protocol. It can also offer premature ovulation risk for an artificially estimated trigger-OPU interval if clinicians prefer to rely on their own expertise. Besides this, ILETIA is also an interpretable tool that provides insights into the importance of clinical variables and how they coactivate to influence the final prediction. These functions aim to enhance clinical decision-making and improve the efficiency of IVF-ET.

ILETIA is principally applicable to PPOS protocol that offers benefits such as a reduced incidence of moderate or severe OHSS [61], oral administration, low cost and clinical convenience, making it one of the most promising COH protocols and a widely adopted choice at our center. Given its significant clinical value, we proposed a public dataset PPOS-DS comprising 8,730 patients receiving PPOS protocol between December 2013 and July 2024 at our center, which can serve as a foundation for future research in IVF-ET.

Transformer [48] is an advanced deep learning architecture based on self-attention mechanism, which has introduced a revolutionary shift in the field of natural language processing. With the exceptional ability to generate diverse representations of the input, it has been recently applied in the field of IVF-ET as well. Nguyen et al. developed an encoder-decoder deformable Transformer network for classifying embryo stage development, which substantially outperformed other current state-of-the-art methods [62]. Wang et al. constructed an AI platform consisting of a Transformer-based network backbone and a self-supervised learning framework for embryo selection [63]. In the investigation of the optimal trigger-OPU interval, introducing Transformer architecture into the method could facilitate the extraction of more latent information from clinical variables. Hence, we designed FTXGB and FTLGB, variants of Transformer specifically for clinical tabular data, which combined FT Transformer architecture [49] and GBT. This inspiration was drawn from previous studies that demonstrated the potential of ensembling deep learning-based models and GBT [54,55]. Our method has been proven to outperform solely trained GBT and FT Transformer, indicating that it retains the potent feature extraction ability of Transformer and the accurate predictive ability of GBT simultaneously, and we further demonstrate that our method has the potential for application beyond PPOS protocol, extending to a broader range of COH protocols.

Although our classifier predicts an interval rank instead of a specific value, this interval rank is still considered valuable for supporting clinical decision-making by clinicians. The median value of each rank (2,130 min for Early, 2,190 min for Mid, 2,250 min for Late) can be approximately used in practice, since each rank spans roughly an hour. This ensures an error margin within 0.5 hours, which is clinically acceptable. Accordingly, ILETIA uses both a classifier (FTXGB) and a regressor (FTLGB) to more effectively aid clinicians in estimating the optimal trigger-OPU interval.

In spite of the encouraging results achieved by ILETIA, we acknowledge that there are several aspects remaining to be improved in future work. First, ILETIA was developed specifically for PPOS protocol due to a lack of sufficient cases of the other protocols that are available for training a Transformer architecture. It's possible to construct a systematic tool for multiple protocols in the future. Second, the clinical variables used in this study didn't include disease-relevant factors (e.g., PCOS) because we unexpectedly found they contributed little to the prediction. A more rational encoding may clarify the relation between disease-relevant factors and the optimal trigger-OPU interval. Finally, we anticipate that training with larger and more diverse real-world datasets could enable ILETIA to predict a broader range of more precise intervals.

Overall, ILETIA is a first-of-its-kind, powerful and interpretable tool that is capable of estimating the trigger-OPU interval. Regarding the clinical application in IVF-ET, we envision that ILETIA can contribute to decision support and comprehension to oocyte retrieval, while also can bridge the gap between clinicians' expertise. In terms of methodology, we hope our framework, Transformer integrated with GBT, will inspire future research that uses clinical tabular data.

# Methods

## Dataset

We propose a dataset named PPOS-DS, which comprises clinical information of 8,730 anonymized patients receiving PPOS protocol conducted at the Ninth People's Hospital affiliated with Shanghai Jiao Tong University School of Medicine between December 2013 and July 2024 (Supplementary Fig. 1a). Using PPOS-DS, we constructed two datasets for the optimal trigger-OPU interval prediction and premature ovulation risk prediction respectively. In this study, the optimal trigger-OPU interval is defined as the actual trigger-OPU interval resulting in an OPU rate ≥0.75. The dataset for the optimal trigger-OPU interval estimation (Dataset A) consists of a training set of 8,383 patients and a test set of 274 patients selected from PPOS-DS (Supplementary Fig. 1b), all with an OPU rate ≥0.75 and a trigger-OPU interval ranging from 2,100 min to 2,280 min. These patients received PPOS protocol between December 2013 and July 2024 (training set: December 2013 to August 2023; test set: September 2023 to July 2024).

For the dataset used for the premature ovulation risk prediction (Supplementary Fig. 1c), the positive cases are premature ovulation patients in PPOS-DS who were identified by transvaginal ultrasound examination or experienced premature ovulation before OPU between July 2015 and July 2023, totally 95 cases. The negative cases (normal patients) were randomly sampled from the training set of Dataset A based on the following criteria: (1) not labeled as premature ovulation, (2) OPU rate =1, (3) 37.8 h≥ trigger-OPU interval ≥36.4 h [7], (4) antral follicle count >5 [64] and (5) E2 ratio>1 [18], totally 400 cases. The criteria ensured that the selected cases had the least premature ovulation risk. Given the imbalance in this subset, we oversampled 55 randomly selected premature ovulation patients by CTGAN and generated 305 additional positive cases. This process resulted in a balanced dataset of 800 cases in total (Dataset B).

## Features and labels

For the optimal trigger-OPU interval prediction, a total of 31 numerical clinical variables were selected as features by experienced clinicians, including 3 health-relevant features, 16 hormone-relevant features, 5 medication-relevant features, 7 follicle-relevant features (Supplementary Table 1). The optimal trigger-OPU intervals of the patients, ranging from 2,100 min to 2,280 min, were used as labels for the regression task. We relabeled the cases in Dataset A by linearly splitting these continuous labels into 3 ranks as follows: Early rank (2,100-2,160 min) for a shorter trigger-OPU interval, Mid rank (2,161-2,219 min) for a mid trigger-OPU interval and Late rank (2,220-2,280 min) for a longer trigger-OPU interval. These discrete labels transformed a regression task into a 3-class classification task. For premature ovulation risk prediction, 2 extra medication-relevant features, MPA dose per day [45], and the artificially estimated interval were used (Supplementary Table 1). Common binary labels were used in this task, where 1 stood for premature ovulation patients (positive), and 0 stood for normal patients (negative).

The clinical variables selected for each of the other 4 COH protocols were not identical (Supplementary Table 1), and different relabeling methods were applied to adapt to the varying data distributions across the 4 protocols. Specifically, for the GnRH antagonist protocol, Early rank covered 2,100-2,159 min, Mid rank covered 2,160-2,189 min and Late rank covered 2,190-2,280 min. For the long protocol, Early rank covered 2,100-2,160 min, Mid rank covered 2,161-2,250 min and Late rank covered 2,251-2,340 min. For the short protocol, Early rank covered 2,070-2,160 min, Mid rank covered 2,161-2,200 min and Late rank covered 2,221-2,310 min. For the mild stimulation protocol, Early rank covered 1,950-2,034 min, Mid rank covered 2,036-2,099 min and Late rank covered 2,100-2,190 min.

## Interval predictors of ILETIA

We designed FTXGB for the 3-class classification task and FTLGB for the regression task as the interval predictors of ILETIA, both of which were adaptations of FT Transformer [49]. FT Transformer (Feature Tokenizer + Transformer Block + MLP) is a simplified version of the Transformer architecture dedicated to tabular data, enabling both numerical and categorical features embeddings. In this research only numerical features were used and MLP was replaced with a GBT model (XGBoost for FTXGB and LightGBM for FTLGB). To summarize, both FTXGB and FTLGB incorporate 3 blocks: Feature Tokenizer, Transformer Block and Gradient-boosted trees (Fig. 1b).

### Feature Tokenizer

Feature Tokenizer is utilized to linearly transformed the numerical features into a high dimension space. For a given feature $x_i$, the corresponding embedding $e_i$ is computed as follows:

$$e_i = b_i + W_i \cdot x_i \quad \in \mathbb{R}^d$$

$$E = \text{stack}[e_1, \ldots, e_k] \quad \in \mathbb{R}^{k \times d}$$

where $b_i$ is the $i$-th feature bias, $W_i \in \mathbb{R}^d$ is $i$-th feature weight, $k$ is the total numerical features and $d$ is the specified dimension to be embedded
Then the [CLS] token is appended to the $E$:

$$E_0 = \text{stack}[[CLS], E]$$

## Transformer Block

After tokenization, $L$ Transformer layers are applied for further feature extraction. PreNorm variant [54] is used to replace original Transformer architecture, in which layer normalization is applied to the input of each sublayer (i.e., multi-head attention sublayer or feed-forward sublayer) of the Transformer layer:

$$Q = \text{LN}(E_l)W_Q, \quad K = \text{LN}(E_l)W_K, \quad V = \text{LN}(E_l)W_V$$

$$\text{Attention}(Q, K, V) = \text{softmax}\left(\frac{QK^T}{\sqrt{d_k}}\right)V$$

$$\text{head}_h = \text{Attention}(Q_h, K_h, V_h)$$

$$\text{MultiHead}(Q, K, V) = \text{Concat}(\text{head}_1, \text{head}_2, \dots, \text{head}_H)W_O$$

$$E'_l = E_l + \text{MultiHead}(Q, K, V)$$

$$E_{l+1} = E'_l + \max(0, \text{LN}(E'_l)W_1 + b_1)W_2 + b_2$$

where LN is layer normalization, $E_l$ is the embedding in $l$-th Transformer layer, $h$ is the specific head and $H$ is the total specified heads. The [CLS] token of the final embedding $E_L$ returned by Transformer Block is used for downstream task through a GBT model.

## Gradient-boosted trees

Different GBTs were chosen as substitutes for MLP in the FT Transformer for 3-class classification and regression because solely trained XGBoost outperformed the other GBT models in the classification task, while LightGBM outperformed the other GBT models in the regression task (Results).
 XGBoost is a supervised ML technique based on the gradient- boosted framework. It focuses on creating an ensemble of sequential decision trees, where each new tree aims to correct the residual errors of its predecessors. In the iteration, higher weights are assigned to misclassified cases. The final output is generated by aggregating the predictions from all the decision trees. This was implemented by Python package xgboost of version 2.1.1 in this research.

LightGBM is also a gradient-boosted framework based on decision trees with higher computational speed and efficiency, which characterized by the following aspects. LightGBM adopts a leaf-wise splitting strategy, which allows the tree to grow by splitting the leaf with the highest loss reduction. Moreover, Gradient-based One-Side Sampling (GOSS), Exclusive Feature Bundling (EFB) and a histogram-based algorithm are introduced to optimize speed, memory usage, and scalability. Python package lightgbm of version 4.5.0 was used in this research.

Model construction

This process is shown in Fig.1b and Supplementary Fig. 2. Initially, a normal FT Transformer (Feature Tokenizer + Transformer Block + MLP) was trained as a 3-class classification task with weighted Cross-Entropy as loss function. The well-trained Feature Tokenizer and Transformer Block were retained with fixed parameters in the subsequent steps and they were used to acquire the embedding $E_L$ from the training set. Next, MLP was substituted with a GBT model (a XGBoost classifier for FTXGB and a LightGBM regressor for FTLGB), which used the [CLS] tokens of $E_L$ as input. In this step, only the GBT was trained, specifically XGBoost classifier with Cross-Entropy as the loss function and LightGBM regressor with Negative Mean Squared Error.

## Oversampling for premature ovulation

In premature ovulation risk prediction, only a limited number of positive cases were collected as premature ovulation is a rare event in IVF-ET, thus the oversampling technique is necessary which can generate new cases of the minority class, and can prevent the model from predicting the majority class. We used CTGAN in this research for oversampling. CTGAN is one of the most prevalent tabular data generators based on Generative Adversarial Networks (GAN) [65]. A mode-specific normalization based on variational Gaussian mixture model is designed to represent columns with complicated distribution, and a conditional generator is designed for balanced sampling to avoid generating only a few categories that frequently appear in the given data. The quality of synthesized data for each feature is evaluated by KSComplement [52] calculated by Kolmogorov-Smirnov statistic, and a higher KSComplement score means higher quality. CTGAN was implemented by Python package sdv of version 1.15.0.

## Model interpretation

We used SHAP, a popular technique for interpreting model predictions, so as to assign each clinical variable an importance value by calculating SHAP values for a particular sample in the test set according to its prediction. This was implemented by the Python package shap of version 0.41.0. For FTXGB and FTLGB, the model-agnostic KernelExplainer that based on the sampling and weighted linear regression was used to calculate SHAP values with background data set to 500. For the premature ovulation risk predictor, the TreeExplainer was used, which was especially designed for tree-based model. The further analysis was implemented by built-in plotting functions that visually presented both local and global interpretation.

## Metrics

Common metrics were used in this research and were computed as follows:
For multi-class classification:

$$\text{Accuracy}_{\text{macro ave}} = \frac{\sum_{i=1}^{C} TP_i}{\sum_{i=1}^{C} TP_i + \sum_{i=1}^{C} FP_i + \sum_{i=1}^{C} FN_i}$$

$$\text{Precision}_{\text{macro ave}} = \frac{1}{C} \sum_{i=1}^{C} \frac{TP_i}{TP_i + FP_i} \qquad \text{Recall}_{\text{macro ave}} = \frac{1}{C} \sum_{i=1}^{C} \frac{TP_i}{TP_i + FN_i}$$

$$\text{F1}_{\text{macro ave}} = \frac{1}{C} \sum_{i=1}^{C} 2 \times \frac{\text{Precision}_i \times \text{Recall}_i}{\text{Precision}_i + \text{Recall}_i}$$

$$\text{Specificity}_{\text{macro ave}} = \frac{1}{C} \sum_{i=1}^{C} \frac{TN_i}{TN_i + FP_i}$$

$$\text{AUROC}_i = \int_0^1 \text{TPR}_i(\text{FPR}_i) \, d\text{FPR}_i \qquad \text{AUPR}_i = \int_0^1 \text{Precision}_i(\text{Recall}_i) \, d\text{Recall}_i$$

$$\text{AUROC}_{\text{macro ave}} = \frac{1}{C} \sum_{i=1}^{C} \text{AUROC}_i \qquad \text{AUPR}_{\text{macro ave}} = \frac{1}{C} \sum_{i=1}^{C} \text{AUPR}_i$$

where $i$ is a specific class and $C$ is the total classes (3 in this research)
For regression:

$$\text{RMSE} = \sqrt{\frac{1}{n} \sum_{k=1}^{n} (y_k - \widehat{y_k})^2} \qquad R^2 = 1 - \frac{\sum_{k=1}^{n}(y_k - \widehat{y_k})^2}{\sum_{k=1}^{n}(y_k - \bar{y})^2}$$

$$r = \frac{\sum_{k=1}^{n}(y_k - \bar{y})(\widehat{y_k} - \bar{\widehat{y}})}{\sqrt{\sum_{k=1}^{n}(y_k - \bar{y})^2 \sum_{k=1}^{n}(\widehat{y_k} - \bar{\widehat{y}})^2}} \qquad \text{MAE} = \frac{1}{n} \sum_{k=1}^{n} |y_k - \widehat{y_k}|$$

where $k$ is a specific sample and $n$ is the total cases
These were simply implemented by scikit-learn [66] and scipy [67].

## Model training and comparison

We didn't do extra data preprocessing for both Dataset A and Dataset B before training except Z-Score Scaler. For both the classification task and the regression task of the optimal trigger-OPU interval, 10-fold cross-validation was performed on the training set of Dataset A firstly, and then the models were trained on Dataset A's training set, and evaluated on Dataset A's test set (Supplementary Fig. 2). The optimal threshold for each class was set to maximize its Youden's J in the 3-class classification task (Early rank: 0.342, Mid rank: 0.653, Late rank: 0.015). Considering that the thresholds may result in clinically unavailable probability, the original outputs are divided by the thresholds and then normalized for probability adjustment. For premature ovulation risk prediction, the test set was a combination of 40 positive cases and 40 negative cases both of which were randomly selected from the total positive or negative cases. Then CTGAN was used to generate 305 new positive cases from the remaining positive cases, which were subsequently combined with the remaining 360 negative cases as the training data. Accordingly, a balanced training set with 720 cases (positive: 360, negative: 360) and test set with 80 cases (positive: 40, negative: 40) were obtained (Supplementary Fig. 1c). Next, a binary classifier was trained (monotonicity constraints were enforced on the feature Trigger-OPU interval).

A series of models were compared with our method. For the optimal trigger-OPU interval 3-class classification, we evaluated 5 traditional ML models including Support Vector Machine (SVM), Random Forest (RF), XGBoost, LightGBM and CatBoost and 6 deep tabular models including FT Transformer, SAINT [68], NODE [69], TabNet [70], STG [71] and MLP. For the optimal trigger-OPU interval regression, we tested 6 models including EN, MLP, SVM, CatBoost, XGBoost and LightGBM. The configuration of each model for a specific task is shown in Supplementary Table 2.

## Data availability

The data that support the results of this study are not publicly available due to privacy concerns, which were used with permission from the Ninth People's Hospital affiliated with Shanghai Jiao Tong University School of Medicine. However, they are available from the corresponding author upon reasonable request and with permission from the Ninth People's Hospital affiliated with Shanghai Jiao Tong University School of Medicine.

# Acknowledgements

This study was supported by grants from the National Key R&D program of China (2023YFF1205103); National Natural Science Foundation of China (81925034, 82441035, 22237005, 82471695, 81925034, 81873856, 32300531), Shanghai Science and Technology Innovation Action Plan, Grant/Award (22Y11906000, 22Y21900800), Shanghai Health and Family Planning Commission (201940287), Shanghai Sailing Program (21YF1422500), the Fund for Excellent Young Scholars of Shanghai Ninth People's Hospital, Shanghai Jiao Tong University School of Medicine (JYYQ013).

# Figures and Tables

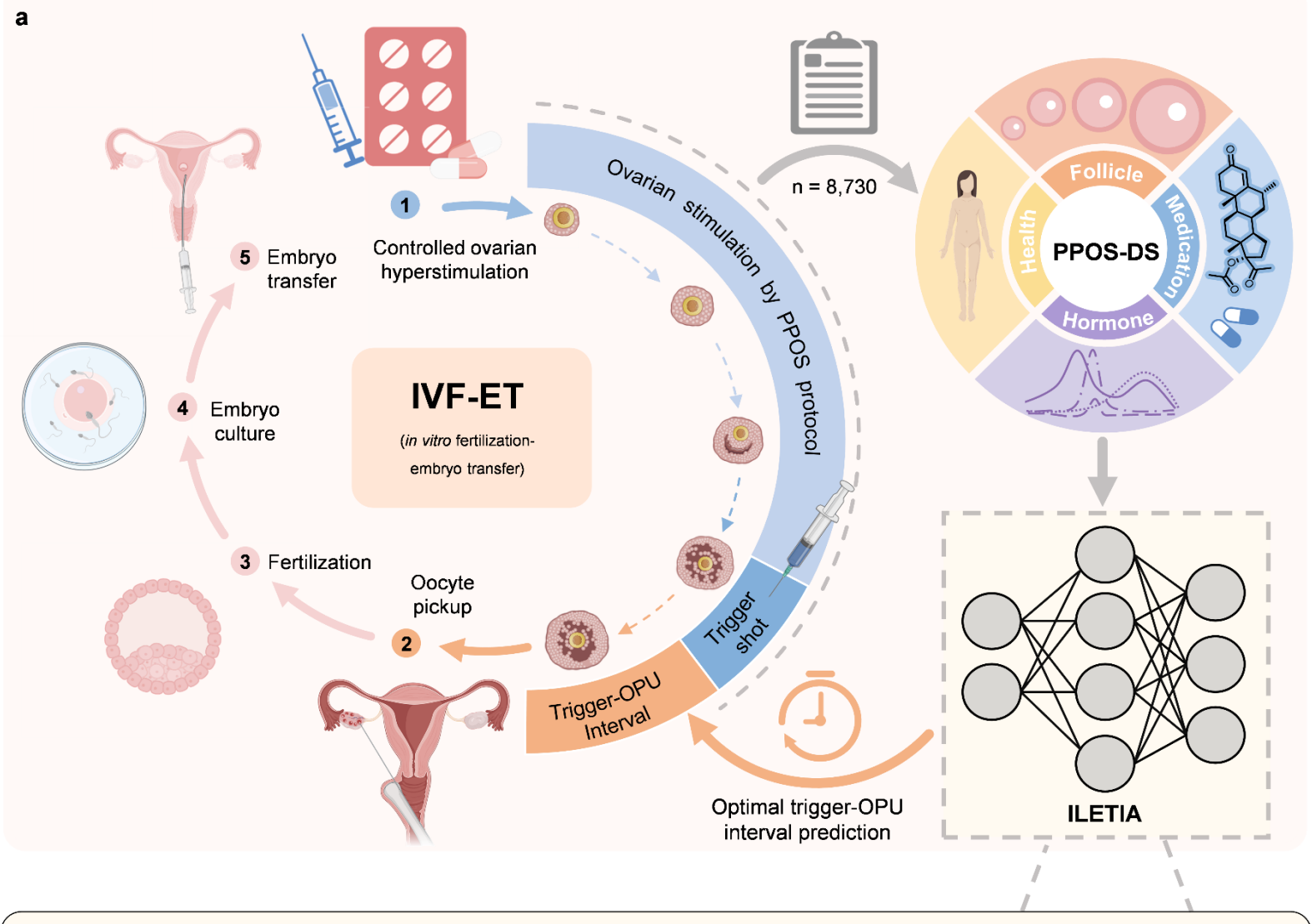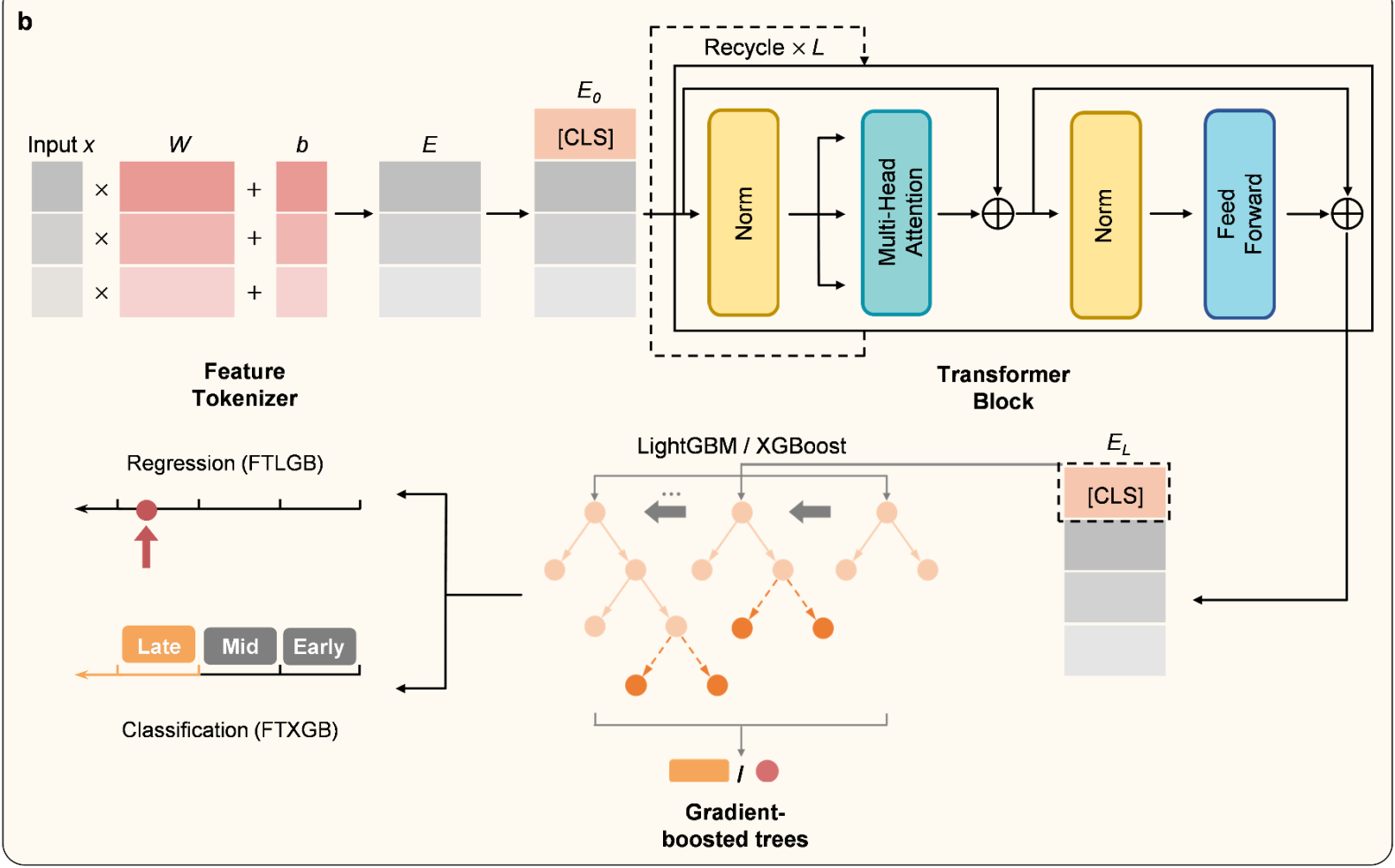

**Fig. 1: General overview of the study.**
**a**, Construction of PPOS-DS and ILETIA. An IVF-ET cycle consists of five key steps. From the start of controlled ovarian hyperstimulation to oocyte pickup, the oocytes gradually grow under the stimulation of medication. PPOS-DS involves clinical information of 8,730 patients during ovarian stimulation using PPOS protocol and around the day of trigger shot, covering 4 aspects: health-relevant features, hormone-relevant features, medication-relevant features and follicle-relevant features. Two models, FTLGB and FTXGB for predicting the optimal trigger-OPU interval are trained with these data, and deployed into a web server, namely ILETIA. Created with [BioRender.com](BioRender.com). **b**, Workflow of FTXGB and FTLGB. The overall architecture comprises 3 blocks: Feature Tokenizer, Transformer Block and gradient-boosted trees (XGBoost for FTXGB and LightGBM for FTLGB). FTXGB operates as a classifier predicting an interval rank with a probability, while FTLGB works as a regressor predicting an interval value. The input data $x$ is passed as a table of numeric features into the model, and linearly transformed by Feature Tokenizer, resulting an embedding $E$. After the tokenization, $L$ layers of Transformer of PreNorm variant are applied for further feature extraction and result the final embedding $E_L$. The [CLS] token of $E_L$ is used as the input of gradient-boosted trees and subsequently the 3-class classification task or the regression task is performed.

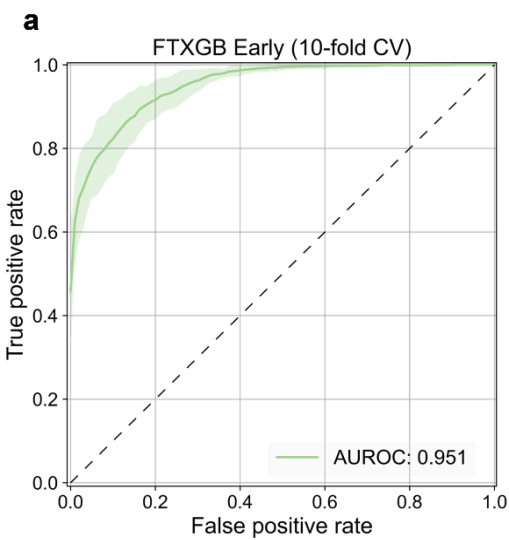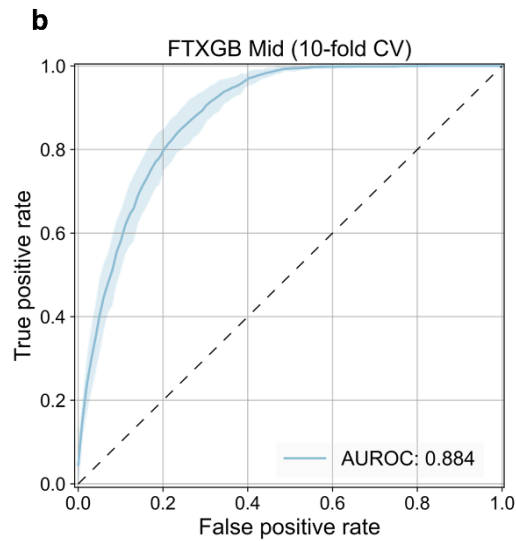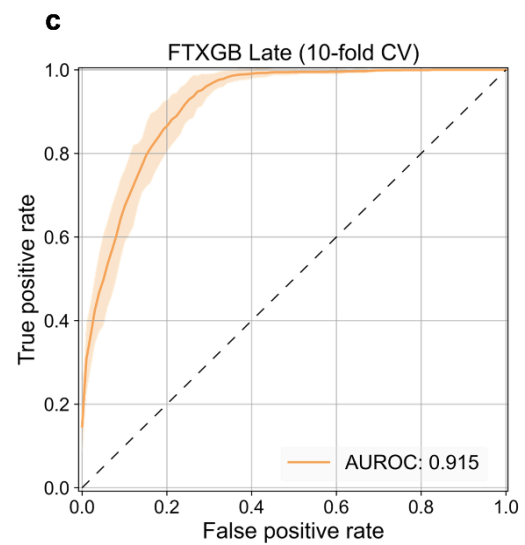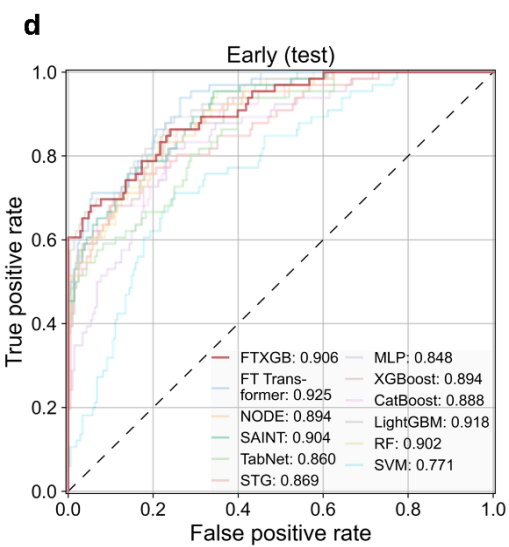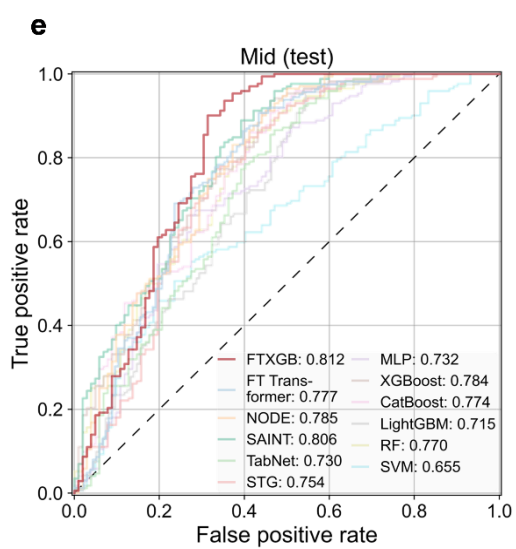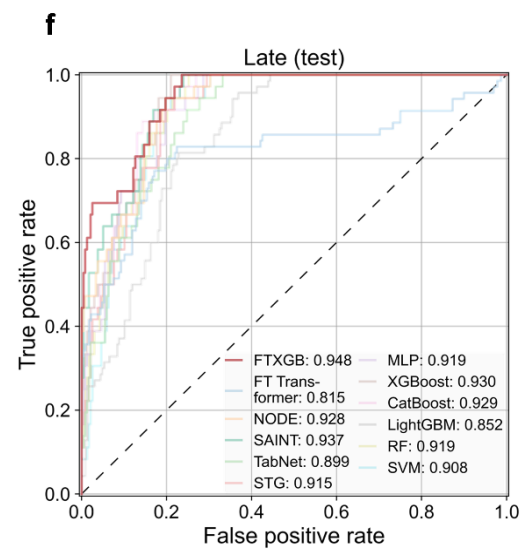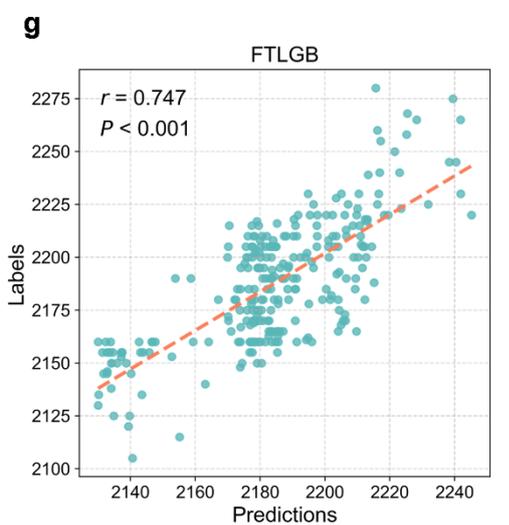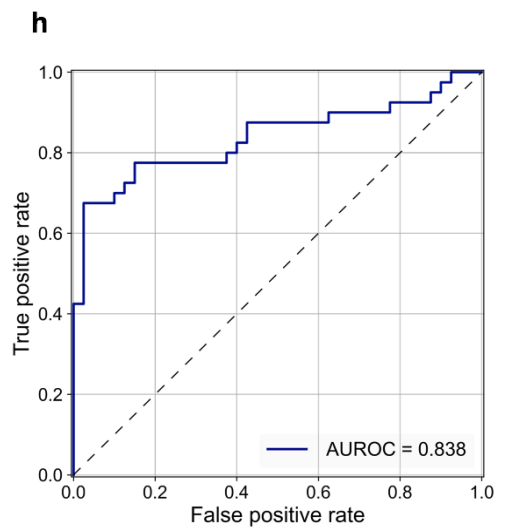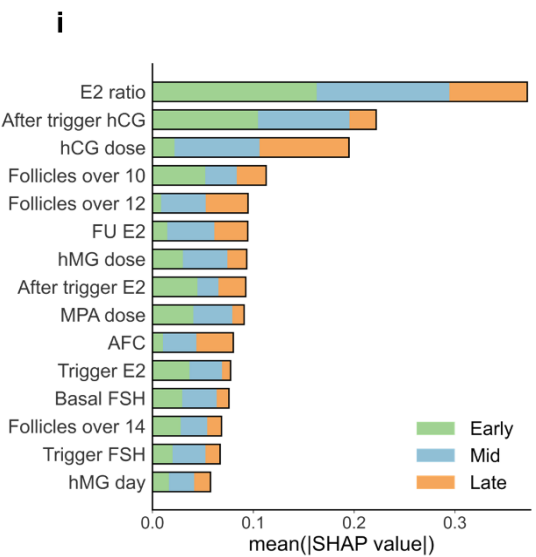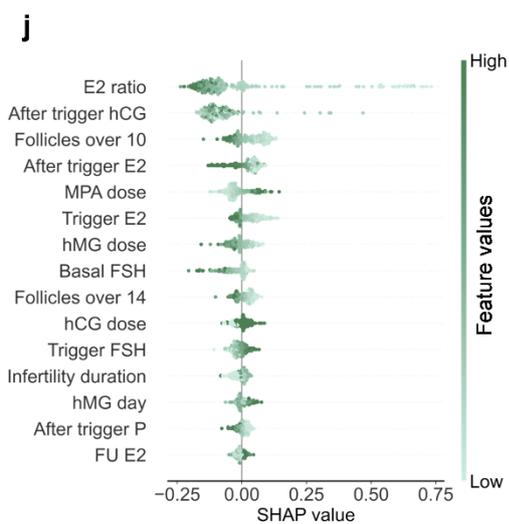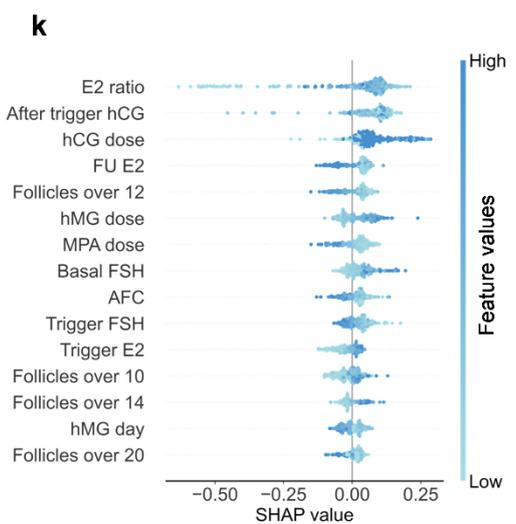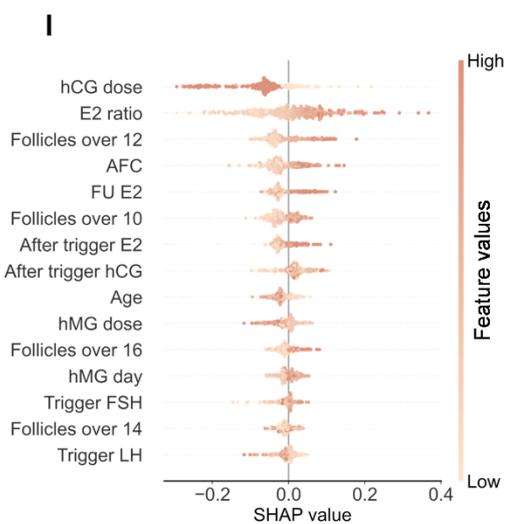

**Fig. 2: Model performance and interpretability.**
**a-c**, Mean ROC for each rank in 10-fold cross-validation of FTXGB. The error bars show the 95% CI of the mean estimate. **d-f**, ROC for each rank on Dataset A's test set across 12 models. **g**, Performance of FTLGB on Dataset A's test set. *r*, Pearson correlation coefficient. **h**, ROC of the premature ovulation risk predictor on Dataset B's test set. **i-l,** Model interpretability of FTXGB by SHAP. **i**, Importance ranking of top-15 clinical variables used in FTXGB based on SHAP values. Different colors represent the contribution of a clinical variable to the predictions within specific interval ranks (green for Early rank, blue for Mid rank and orange for Late rank). **j-l**, SHAP beeswarm plots of top-15 most important clinical variables for each interval rank: Early (**j**), Mid (**k**), Late (**l**). For each interval rank, dots or samples with a deeper color represent higher feature values and those with a lighter color represent lower feature values. A positive SHAP value indicates an increased probability of predicting the specific interval rank and vice versa.

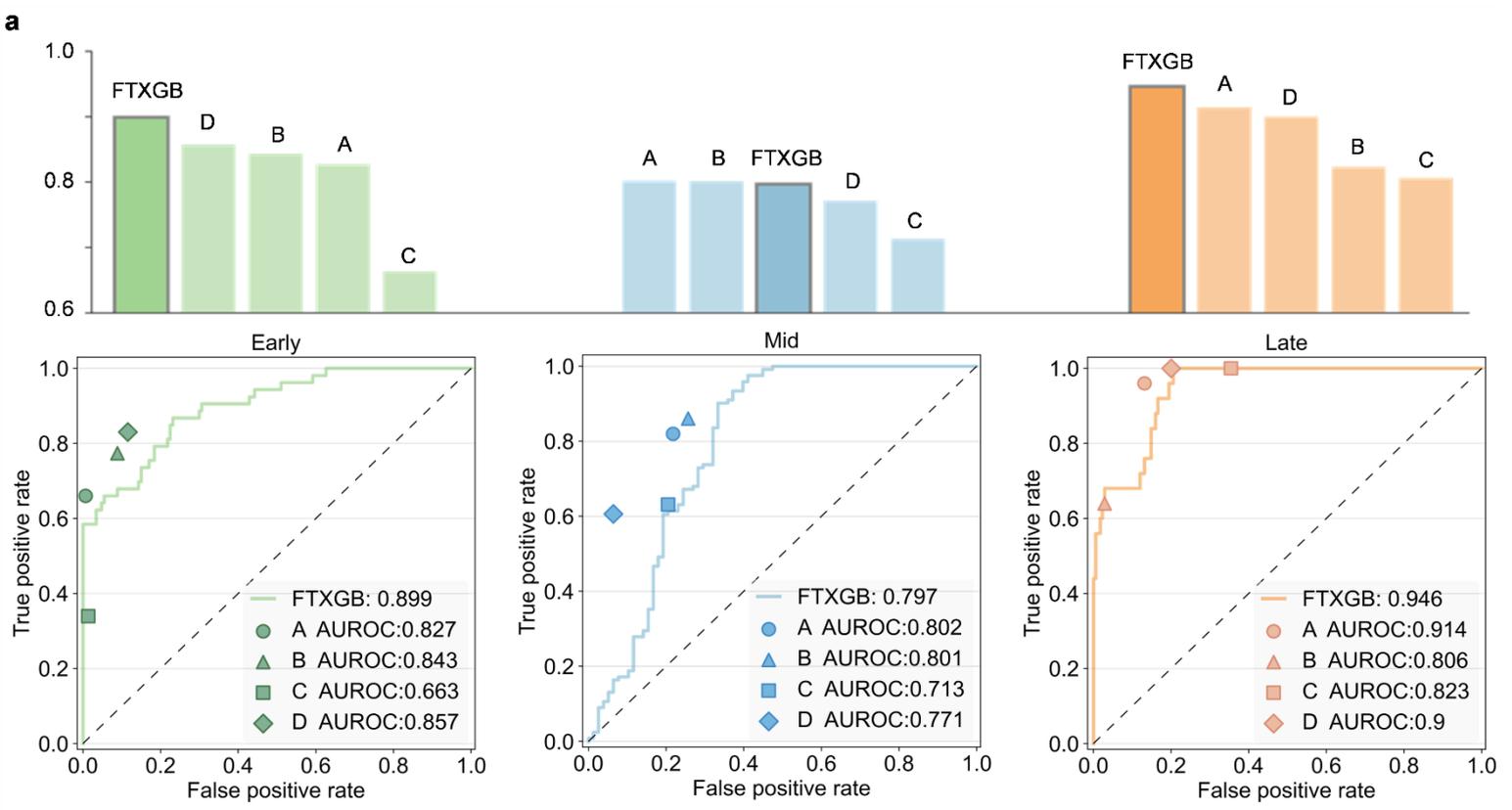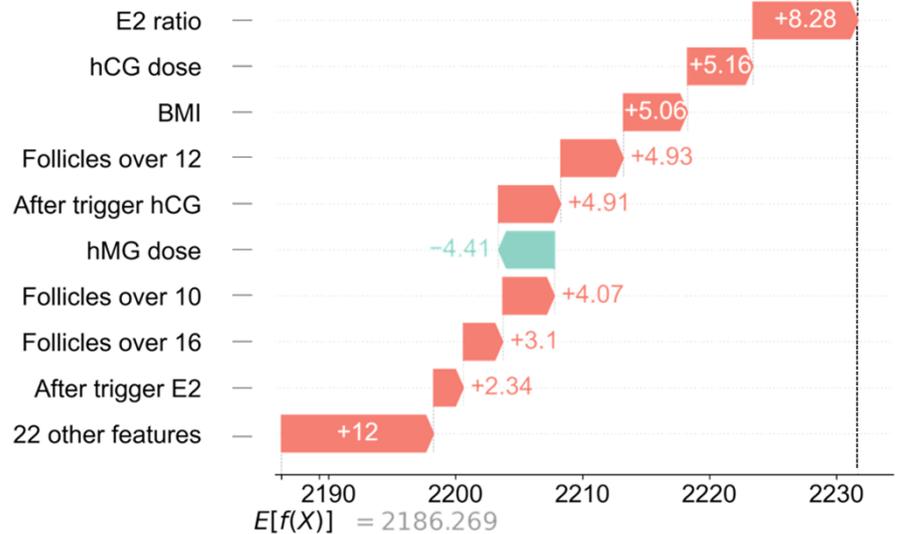

**Fig. 3: The potential of ILETIA to assist in clinical decision-making.**
**a**, Comparison of FTXGB with 4 specialists across 3 ranks. The dataset used in the comparison are 200 samples randomly selected from Dataset A's test set. For each rank, the bar plot above illustrates the ranking of AUROCs of FTXGB and the 4 specialists, while the bottom figure shows ROC of FTXGB alongside the prediction performances of the 4 specialists. **b-d**, Analysis of one case of the second oocyte retrieval using ILETIA. The output includes the clinical variables used (**b**), the predictions (**c**) of both FTLGB (interval value) and FTXGB (interval rank), and the waterfall plot of FTLGB's interpretation (**d**). The base value $E[f(X)]$ represents the average prediction of the regressor considering no features. It is adjusted to the final prediction by the contributions of clinical variables. Variables with orange bars increase the prediction, while those with green bars decrease it, and they are ranked by their contribution in the left column.

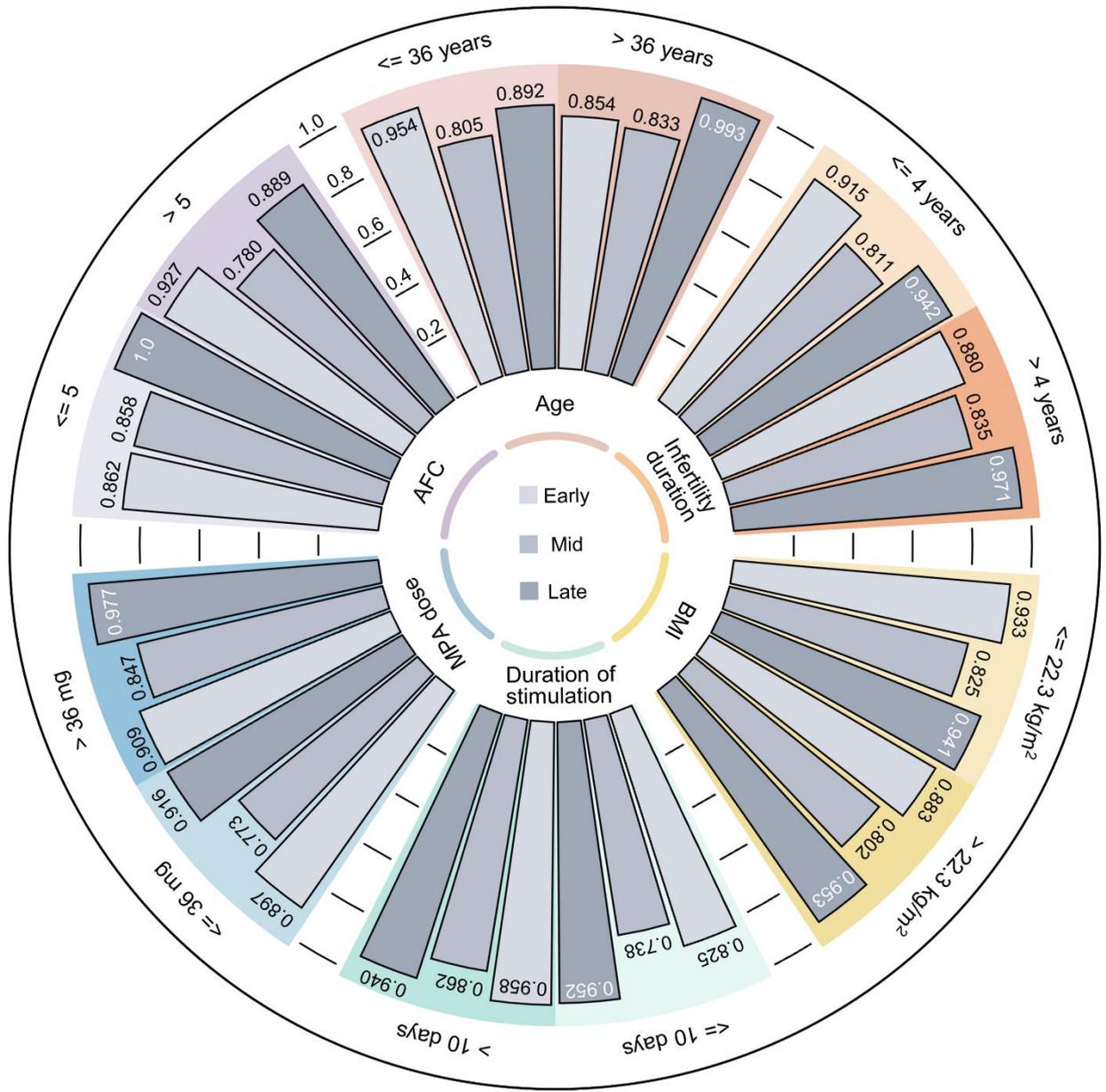
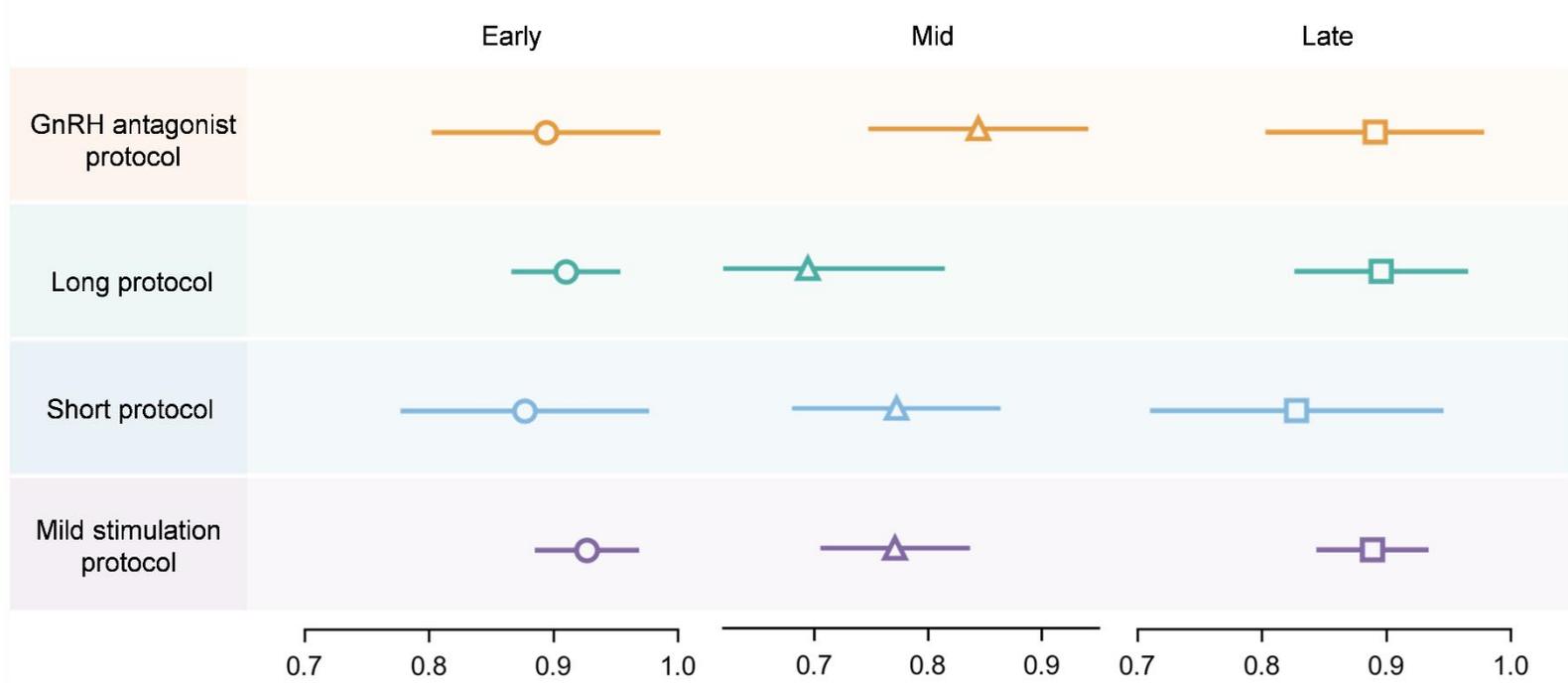

**Fig. 4: Robustness of FTXGB across different subgroups and its general utility to other COH protocols.**

**a,** Subgroup analysis of 6 clinical variables. For each clinical variable, Dataset A's test set is split into two subgroups according to a specified condition and then evaluated by FTXGB respectively, with the results are calculated as AUROCs of the 3 ranks. The inner circle displays the names of the 6 clinical variables, with each corresponding to a specific cell. The light and dark colors represent different subgroups respectively. In each subgroup, the grey bars represent AUROCs of each rank in sequence. **b,** AUROCs of the 3 ranks in 10-fold cross-validation for GnRH antagonist protocol, long protocol, short protocol and mild stimulation protocol. The circle, triangle and square represent Early, Mid and Late rank respectively. The error bars show the 95% CI of the mean estimate.

| 10-fold cross-validation | | | | | |
|---|---|---|---|---|---|
| | **SVM** | **RF** | **LightGBM** | **CatBoost** | **XGBoost** |
| ACC | 0.633 | 0.743 | 0.748 | 0.750 | 0.759 |
| PR | 0.629 | 0.749 | 0.750 | 0.752 | 0.765 |
| RE | 0.626 | 0.750 | 0.752 | 0.756 | 0.749 |
| SP | 0.809 | 0.868 | 0.870 | 0.872 | 0.871 |
| F1 | 0.624 | 0.746 | 0.749 | 0.752 | 0.755 |
| AUROC | 0.814 | 0.905 | 0.911 | 0.913 | 0.911 |
| AUPR | 0.664 | 0.822 | 0.833 | 0.840 | 0.835 |

| | **MLP** | **STG** | **TabNet** | **SAINT** | **NODE** | **FT Transformer** | **FTXGB (Ours)** |
|---|---|---|---|---|---|---|---|
| ACC | 0.725 | 0.731 | 0.714 | 0.750 | 0.755 | 0.758 | **0.766** |
| PR | 0.730 | 0.736 | 0.721 | 0.759 | 0.763 | 0.759 | **0.773** |
| RE | 0.719 | 0.724 | 0.720 | 0.739 | 0.745 | **0.765** | 0.758 |
| SP | 0.855 | 0.858 | 0.853 | 0.866 | 0.869 | 0.876 | **0.876** |
| F1 | 0.722 | 0.728 | 0.716 | 0.746 | 0.752 | 0.760 | **0.764** |
| AUROC | 0.889 | 0.894 | 0.882 | 0.908 | 0.909 | 0.917 | **0.917** |
| AUPR | 0.799 | 0.806 | 0.780 | 0.831 | 0.835 | **0.848** | 0.847 |

| test set | | | | | |
|---|---|---|---|---|---|
| | **SVM** | **RF** | **LightGBM** | **CatBoost** | **XGBoost** |
| ACC | 0.606 | 0.759 | 0.734 | 0.737 | 0.781 |
| PR | 0.568 | 0.728 | 0.679 | 0.694 | 0.779 |
| RE | 0.670 | 0.675 | 0.708 | 0.654 | 0.657 |
| SP | 0.789 | 0.826 | 0.834 | 0.817 | 0.827 |
| F1 | 0.588 | 0.694 | 0.692 | 0.671 | 0.699 |
| AUROC | 0.778 | 0.864 | 0.864 | 0.864 | 0.869 |
| AUPR | 0.594 | 0.747 | 0.763 | 0.763 | 0.770 |

|        | MLP   | STG   | TabNet | SAINT | NODE  | FT Transformer | FTXGB (Ours) |
|--------|-------|-------|--------|-------|-------|----------------|--------------|
| ACC    | 0.682 | 0.730 | 0.734  | 0.785 | 0.792 | 0.796          | **0.821**    |
| PR     | 0.629 | 0.673 | 0.690  | 0.766 | **0.874** | 0.804      | 0.813        |
| RE     | 0.680 | 0.689 | 0.626  | 0.702 | 0.642 | 0.693          | **0.759**    |
| SP     | 0.816 | 0.828 | 0.806  | 0.841 | 0.824 | 0.843          | **0.869**    |
| F1     | 0.648 | 0.680 | 0.650  | 0.727 | 0.703 | 0.734          | **0.782**    |
| AUROC  | 0.833 | 0.846 | 0.830  | 0.883 | 0.869 | 0.889          | **0.889**    |
| AUPR   | 0.702 | 0.723 | 0.694  | 0.798 | 0.776 | 0.817          | **0.819**    |

**Table 1: Performance metrics in 10-fold cross validation and on the test set for 3-class classification task of PPOS protocol.**
The first two tables show the results of models performing 10-fold cross-validation on Dataset A's training set, while the last two tables show those on Dataset A's test set. For each situation, the results are divided into two sections: the first displays the results of traditional machine learning models and the second focuses on deep tabular models. For all metrics, best results are shown in bold. FTXGB outperforms the other methods in most metrics in both the cross-validation and on the test set. ACC: macro-average accuracy, PR: macro-average precision, RE: macro-average recall, SP: macro-average specificity, F1: macro-average F1 score, AUROC: macro-average area under ROC curve, AUPR: macro-average area under precision-recall curve.

|  | 10-fold cross-validation | | | | | | |
|---|---|---|---|---|---|---|---|
|  | EN | MLP | RF | XGBoost | CatBoost | LightGBM | FTLGB (Ours) |
| RMSE | 35.53 | 52.44 | 26.82 | 27.16 | 27.02 | 27.04 | **24.85** |
| MAE | 26.41 | 27.73 | 21.03 | 21.37 | 21.26 | 21.29 | **19.56** |
| r | 0.531 | 0.487 | 0.754 | 0.747 | 0.750 | 0.750 | **0.793** |

|  | test set | | | | | |
|---|---|---|---|---|---|---|
|  | EN | RF | XGBoost | CatBoost | LightGBM | FTLGB (Ours) |
| RMSE | 25.87 | 21.12 | 21.65 | 21.72 | 20.78 | **20.01** |
| MAE | 20.21 | 17.29 | 17.69 | 17.40 | 17.07 | **16.70** |
| r | 0.573 | 0.718 | 0.703 | 0.692 | 0.719 | **0.747** |

**Table 2: Performance metrics in 10-fold cross validation and on the test set for regression task of PPOS protocol.**
The upper table shows the results of models performing 10-fold cross-validation on Dataset A's training set, while the lower one shows those on Dataset A's test set. For all metrics, best results are shown in bold. FTLGB outperforms the other methods in all 3 metrics. MLP performed poorly in 10 cross-validation, thus we did not further evaluate it on Dataset A's test set. RMSE: Root Mean Squared Error, MAE: Mean Absolute Error, r: Pearson correlation coefficient.

|  | SVM | RF | LightGBM | CatBoost | XGBoost |
|---|---|---|---|---|---|
| **GnRH antagonist protocol** | | | | | |
| AUROC | 0.753 | 0.838 | 0.857 | 0.859 | 0.860 |
| AUPR | 0.613 | 0.714 | 0.750 | 0.751 | 0.751 |
| **Long protocol** | | | | | |
| AUROC | 0.753 | 0.837 | 0.841 | 0.843 | 0.844 |
| AUPR | 0.580 | 0.708 | 0.723 | 0.727 | 0.726 |
| **Short protocol** | | | | | |
| AUROC | 0.684 | 0.819 | 0.830 | 0.826 | 0.837 |
| AUPR | 0.509 | 0.664 | 0.675 | 0.675 | 0.684 |
| **Mild stimulation protocol** | | | | | |
| AUROC | 0.811 | 0.857 | 0.863 | 0.861 | 0.864 |
| AUPR | 0.691 | 0.755 | 0.764 | 0.761 | 0.764 |

|  | MLP | STG | TabNet | SAINT | NODE | FT Transformer | FTXGB (Ours) |
|---|---|---|---|---|---|---|---|
| **GnRH antagonist protocol** | | | | | | | |
| AUROC | 0.821 | 0.827 | 0.766 | 0.861 | 0.862 | 0.870 | 0.876 |
| AUPR | 0.688 | 0.699 | 0.613 | 0.747 | 0.755 | 0.758 | 0.769 |
| **Long protocol** | | | | | | | |
| AUROC | 0.805 | 0.789 | 0.783 | 0.831 | 0.830 | 0.833 | 0.834 |
| AUPR | 0.653 | 0.634 | 0.613 | 0.709 | 0.701 | 0.711 | 0.713 |
| **Short protocol** | | | | | | | |
| AUROC | 0.770 | 0.796 | 0.767 | 0.819 | 0.818 | 0.821 | 0.826 |
| AUPR | 0.602 | 0.630 | 0.592 | 0.661 | 0.656 | 0.665 | 0.669 |
| **Mild stimulation protocol** | | | | | | | |
| AUROC | 0.837 | 0.842 | 0.816 | 0.856 | 0.862 | 0.860 | 0.862 |
| AUPR | 0.726 | 0.738 | 0.690 | 0.754 | 0.763 | 0.760 | 0.762 |

**Table 3: Performance metrics for 3-class classification task of GnRH antagonist protocol, long protocol, short protocol and mild stimulation protocol.**
For each COH protocol, 10-fold cross-validation is performed on the corresponding dataset. The upper table shows the results of traditional machine learning models and the lower one displays those of deep tabular models. Our methods outperform the other deep tabular models across all 4 COH protocols. More details in Supplementary Table 6. AUROC, macro-average area under ROC curve. AUPR, macro-average area under precision-recall curve.

# Supplementary materials

**Table S1: Features used as input for model training in the study.**
Features listed under 'all 5 protocols' in the last column indicate that these were used in the optimal trigger-OPU prediction for patients receiving PPOS protocol, GnRH antagonist protocol, long protocol, short protocol and mild stimulation protocol. Specifically, 2 features were used only in premature ovulation risk prediction in PPOS protocol: hMG dose per day and Trigger-OPU interval estimated by clinicians.

| Feature name | Unit | Definition | Protocol using the feature |
|---|---|---|---|
| Age | year | age on the oocyte pickup day | all 5 protocols: |
| Infertility duration | year | the duration of infertility. | all 5 protocols |
| BMI | kg/m$^2$ | weight / height$^2$ | all 5 protocols |
| Duration of stimulation | day | days in a specific controlled ovarian hyperstimulation protocol cycle before the trigger shot | all 5 protocols |
| Basal FSH | mIU/mL | the level of follicle-stimulating hormone on the day 2 or 3 of the menstrual cycle | all 5 protocols |
| Basal LH | mIU/mL | the level of luteinizing hormone on the day 2 or 3 of the menstrual cycle | all 5 protocols |
| Basal E2 | pg/mL | the level of estradiol on the day 2 or 3 of the menstrual cycle | all 5 protocols |
| Basal P | ng/mL | the level of progesterone on the day 2 or 3 of the menstrual cycle | all 5 protocols |
| FU LH | mIU/mL | the level of luteinizing hormone on the day of follow-up | all 5 protocols |
| FU E2 | pg/mL | the level of estradiol on the day of follow-up | all 5 protocols |
| Trigger FSH | mIU/mL | the level of follicle-stimulating hormone on the trigger day | all 5 protocols |
| Trigger LH | mIU/mL | the level of luteinizing hormone on the trigger day | all 5 protocols |

| Name | Unit | Description | Protocol |
|---|---|---|---|
| Trigger E2 | pg/mL | the level of estradiol on the trigger day | all 5 protocols |
| Trigger P | ng/mL | the level of progesterone on the trigger day | all 5 protocols |
| After trigger hCG | mIU/mL | the level of human chorionic gonadotropin after the trigger day | all 5 protocols |
| After trigger FSH | mIU/mL | the level of follicle-stimulating hormone after the trigger day | all 5 protocols |
| After trigger LH | mIU/mL | the level of luteinizing hormone after the trigger day | all 5 protocols |
| After trigger E2 | pg/mL | the level of estradiol after the trigger day | all 5 protocols |
| E2 ratio | | After trigger E2 / Trigger E2 | all 5 protocols |
| After trigger P | ng/mL | the level of progesterone after the trigger day | all 5 protocols |
| hMG day | day | days of using human menopausal gonadotropin | PPOS protocol GnRH antagonist protocol long protocol short protocol |
| Dabigatran day | day | days of using Dabigatran | mild stimulation |
| hMG dose | IU | total dose of human menopausal gonadotropin used | PPOS protocol GnRH antagonist protocol long protocol short protocol |
| hMG dose per day | IU | dose of human menopausal gonadotropin used in a single day | PPOS protocol (premature ovulation risk only) |
| hCG dose | IU | total dose of human chorionic gonadotropin used | PPOS protocol GnRH antagonist protocol long protocol short protocol |
| Clomiphene dose | mg | total dose of Clomiphene used | mild stimulation |
| Leuprorelin acetate dose | mg | total dose of Leuprorelin acetate used | long protocol |
| Dabigatran dose | mg | total dose of Dabigatran used | mild stimulation |
| MPA dose | mg | total dose of medroxyprogesterone acetate | PPOS protocol |

|  |  | used |  |
|---|---|---|---|
| AFC |  | antral follicle count on the day 2 or 3 of the menstrual cycle | all 5 protocols |
| Follicles over 10 |  | the number of follicles with diameter over 10mm on the trigger day | all 5 protocols |
| Follicles over 12 |  | the number of follicles with diameter over 12mm on the trigger day | all 5 protocols |
| Follicles over 14 |  | the number of follicles with diameter over 14mm on the trigger day | all 5 protocols |
| Follicles over 16 |  | the number of follicles with diameter over 16mm on the trigger day | all 5 protocols |
| Follicles over 18 |  | the number of follicles with diameter over 18mm on the trigger day | all 5 protocols |
| Follicles over 20 |  | the number of follicles with diameter over 20mm on the trigger day | all 5 protocols |
| Estimated interval | hour | the trigger-OPU interval estimated by clinicial | PPOS protocol (only used in premature ovulation) |

**Table S2: Model configurations for trigger-OPU prediction on Dataset A's test set and premature ovulation prediction on Dataset B's test set.**
The same model might be used for both regression and 3-class classification tasks with different hyperparameters. The hyperparameters not mentioned here were set to their default values.

| Model | Task | Configuration |
|---|---|---|
| Elastic Net | regression of the optimal trigger-OPU interval | alpha: 0.1<br>l1 ratio: 1.0 |
| Random Forest | regression of the optimal trigger-OPU interval | max depth: 10<br>min samples leaf: 8<br>min samples split: 5<br>n estimators: 500<br>sample weight: 0.65 when label $\in$ [2125, 2225] |
| CatBoost | regression of the optimal trigger-OPU interval | l2 leaf reg: 1<br>depth: 4<br>learning rate: 0.03<br>early stopping rounds: 50 |
| LightGBM | regression of the optimal trigger-OPU interval | learning rate: 0.1<br>num leaves: 30<br>max depth: 7<br>feature fraction: 1<br>bagging fraction: 1<br>bagging freq: 3<br>stopping rounds: 50 |
| XGBoost | regression of the optimal trigger-OPU interval | max depth: 3<br>min child weight: 6<br>learning rate: 0.13<br>colsample bytree: 0.6<br>subsample: 0.7<br>gamma: 0.1<br>reg lambda: 1.0<br>reg alpha: 0.0<br>early stopping rounds: 70 |

| Model | Task | Hyperparameters |
|---|---|---|
| FTLGB | regression of the optimal trigger-OPU interval | FT Transformer part:<br>  learning rate: 1e-4<br>  weight decay: 1e-4<br>  dim: 32<br>  dim out: 3<br>  depth: 4<br>  heads: 8<br>  attn dropout: 0.1<br>  ff dropout: 0.1<br>  early stopping rounds: 50<br>LightGBM part:<br>  learning rate: 0.1<br>  num leaves: 30<br>  max depth: 3<br>  feature fraction: 0.7<br>  bagging fraction: 0.85<br>  bagging freq: 3<br>  n estimators: 100<br>  stopping rounds: 50 |
| Support Vector Machine | 3-class classification of the optimal trigger-OPU interval | C: 7<br>gamma: 0.0001<br>kernel: rbf |
| Random Forest | 3-class classification of the optimal trigger-OPU interval | max depth: 10<br>min samples leaf: 1<br>min samples split: 5<br>n estimators: 1200 |
| LightGBM | 3-class classification of the optimal trigger-OPU interval | learning rate: 0.1<br>num leaves: 50<br>max depth: 5<br>feature fraction: 0.7<br>bagging fraction: 0.85<br>bagging freq: 3<br>stopping rounds: 50 |
| CatBoost | 3-class classification of the optimal trigger-OPU interval | l2 leaf reg: 9<br>depth: 5<br>learning rate: 0.1<br>early stopping rounds: 50 |
| XGBoost | 3-class classification of the optimal trigger-OPU interval | max depth: 3<br>min child weight: 9<br>learning rate: 0.13<br>colsample bytree: 0.9<br>subsample: 0.7<br>gamma: 0.3<br>reg lambda: 1.5 |

| | | reg alpha: 0.5 |
| --- | --- | --- |
| | | early stopping rounds: 50 |
| Multilayer Perceptron | 3-class classification of the optimal trigger-OPU interval | activation: tanh |
| | | alpha: 0.05 |
| | | early stopping: True |
| | | hidden layer sizes: (128, 128) |
| | | learning rate: adaptive |
| | | solver: adam |
| | | validation fraction: 0.1 |
| STG | 3-class classification of the optimal trigger-OPU interval | learning rate: 1.48e-3 |
| | | lam: 1.87e-3 |
| | | hidden dims: [ 500, 400, 20] |
| | | early stopping rounds: 70 |
| TabNet | 3-class classification of the optimal trigger-OPU interval | n d: 48 |
| | | n a: 8 |
| | | n steps: 4 |
| | | gamma: 2.0 |
| | | n independent: 7 |
| | | n shared: 2 |
| | | learning rate: 0.005 |
| SAINT | 3-class classification of the optimal trigger-OPU interval | dim: 81 |
| | | depth: 3 |
| | | heads: 4 |
| | | dropout: 0.4 |
| | | early stopping rounds: 50 |
| NODE | 3-class classification of the optimal trigger-OPU interval | num_layers: 2 |
| | | total_tree_count: 1024 |
| | | tree_depth: 8 |
| | | tree_output_dim: 3 |
| | | early stopping rounds: 50 |
| FT Transformer | 3-class classification of the optimal trigger-OPU interval | learning rate: 1e-4 |
| | | weight decay: 1e-4 |
| | | dim: 32 |
| | | dim out: 3 |
| | | depth: 4 |
| | | heads: 8 |
| | | attn dropout: 0.1 |
| | | ff dropout: 0.1 |
| | | early stopping rounds: 50 |
| FTXGB | 3-class classification of the optimal trigger-OPU interval | FT Transformer part:<br>　learning rate: 1e-4<br>　weight decay: 1e-4<br>　dim: 32<br>　dim out: 3<br>　depth: 4<br>　heads: 8<br>　attn dropout: 0.1<br>　ff dropout: 0.1<br>　early stopping rounds: 50<br>XGBoost part:<br>　max depth: 3<br>　min child weight: 9<br>　learning rate: 0.13<br>　colsample bytree: 0.9<br>　subsample: 0.7<br>　gamma: 0.3<br>　reg lambda: 1.5<br>　reg alpha: 0.5<br>　early stopping rounds: 70 |
| XGBoost | binary classification of premature ovulation | max depth: 7<br>min child weight: 2<br>learning rate: 0.13<br>colsample bytree: 0.8<br>subsample: 0.8<br>gamma: 0.1<br>reg lambda: 0.6<br>reg alpha: 0.2<br>monotone constraints: Trigger-OPU - interval :1<br>early stopping rounds: 50 |

**Table S3: Quality of positive samples generated by CTGAN across 32 features used for premature ovulation risk prediction.**

305 new positive samples were generated by CTGAN based on 55 original positive samples as part of Dataset B's training set. The metric used to evaluate the quality of the new samples for each feature is KSComplement, which is calculated by Kolmogorov-Smirnov statistic. A higher score indicates higher quality, and vice versa. The average score is 0.800.

| Column | Metric | Score |
| --- | --- | --- |
| Age | KSComplement | 0.876602 |
| Infertility duration | KSComplement | 0.938599 |
| BMI | KSComplement | 0.872727 |
| Duration of stimulation | KSComplement | 0.754694 |
| Basal FSH | KSComplement | 0.853353 |
| Basal LH | KSComplement | 0.776751 |
| Basal E2 | KSComplement | 0.897466 |
| Basal P | KSComplement | 0.723994 |
| FU LH | KSComplement | 0.864382 |
| FU E2 | KSComplement | 0.740984 |
| Trigger FSH | KSComplement | 0.767511 |
| Trigger LH | KSComplement | 0.750224 |
| Trigger E2 | KSComplement | 0.794933 |
| Trigger P | KSComplement | 0.790164 |
| After trigger hCG | KSComplement | 0.863189 |
| After trigger FSH | KSComplement | 0.662295 |
| After trigger LH | KSComplement | 0.771982 |
| After trigger E2 | KSComplement | 0.761848 |
| E2 ratio | KSComplement | 0.5538 |
| After trigger P | KSComplement | 0.783607 |
| hMG stimulation | KSComplement | 0.615201 |
| hMG dose | KSComplement | 0.622653 |
| hCG dose | KSComplement | 0.904322 |
| MPA dose | KSComplement | 0.85842 |
| AFC | KSComplement | 0.754694 |
| Follicles over 10 | KSComplement | 0.82623 |
| Follicles over 12 | KSComplement | 0.830402 |
| Follicles over 14 | KSComplement | 0.833979 |
| Follicles over 16 | KSComplement | 0.831595 |
| Follicles over 18 | KSComplement | 0.939195 |
| Follicles over 20 | KSComplement | 0.908197 |
| Estimated interval | KSComplement | 0.8769 |

**Table S4: 33 cases of the second oocyte retrieval.**
If conditions permit, when a patient's OPU rate is low in the initial OPU, the initial trigger-OPU interval could be extended and the second OPU is performed to increase OPU rate. The second intervals are shown with the interval values and their corresponding ranks (Early: 2,100-2,160 min, Mid: 2,161-2,219 min, Late: 2,220-2,280 min), and those longer than 2,280min are also labeled as Late. Cases with a grey background are consistent with the corresponding second interval. Cases with a yellow background are closer to the second interval than the initial. Cases with no background are not aligned with the corresponding second interval. The input and output of the specific case with a red border are shown in Fig. 3**b-d**.

| Initial interval (min) | Initial OPU rate | Second interval (min) | Second OPU rate | Result of FTXGB (Early, Mid, Late) |
|---|---|---|---|---|
| 2190 | 0.143 | 2250 (Late) | 1.0 | 0.002, 0.052, 0.946 (Late) |
| 2190 | 0.167 | 2250 (Late) | 0.5 | 0.228, 0.524, 0.247 (Mid) |
| 2160 | 0.385 | 2220 (Late) | 1.0 | 0.001, 0.049, 0.95 (Late) |
| 2190 | 0.0 | 2310 (Late) | 0.667 | 0.052, 0.659, 0.289 (Mid) |
| 2187 | 0.0 | 2250 (Late) | 1.0 | 0.301, 0.511, 0.188 (Mid) |
| 2178 | 0.182 | 2238 (Late) | 0.545 | 0.001, 0.031, 0.969 (Late) |
| 2178 | 0.0 | 2250 (Late) | 1.0 | 0.138, 0.683, 0.179 (Mid) |
| 2220 | 0.123 | 2310 (Late) | 0.714 | 0.0, 0.0, 1.0 (Late) |
| 2190 | 0.167 | 2310 (Late) | 0.75 | 0.0, 0.0, 1.0 (Late) |
| 2220 | 0.0 | 2310 (Late) | 0.889 | 0.112, 0.333, 0.555 (Late) |
| 2250 | 0.0 | 2310 (Late) | 0.429 | 0.001, 0.021, 0.979 (Late) |
| 2160 | 0.25 | 2280 (Late) | 1.0 | 0.115, 0.687, 0.198 (Mid) |
| 2190 | 0.2 | 2310 (Late) | 0.333 | 0.082, 0.74, 0.178 (Mid) |
| 2220 | 0.444 | 2310 (Late) | 1.0 | 0.0, 0.0, 0.999 (Late) |
| 2160 | 0.25 | 2232 (Late) | 0.75 | 0.04, 0.712, 0.248 (Mid) |
| 2226 | 0.0 | 2286 (Late) | 0.6 | 0.001, 0.022, 0.978 (Late) |
| 2100 | 0.0 | 2160 (Early) | 0.5 | 0.499, 0.328, 0.174 (Early) |
| 2130 | 0.0 | 2280 (Late) | 0.75 | 0.129, 0.686, 0.185 (Mid) |
| 2250 | 0.0 | 2310 (Late) | 0.6 | 0.001, 0.069, 0.93 (Late) |
| 2160 | 0.333 | 2220 (Late) | 1.0 | 0.017, 0.451, 0.532 (Late) |
| 2160 | 0.0 | 2220 (Late) | 1.0 | 0.001, 0.028, 0.971 (Late) |
| 2160 | 0.471 | 2220 (Late) | 1.0 | 0.001, 0.017, 0.983 (Late) |
| 2190 | 0.25 | 2250 (Late) | 0.667 | 0.0, 0.002, 0.997 (Late) |
| 2166 | 0.0 | 2250 (Late) | 0.4 | 0.13, 0.668, 0.201 (Mid) |
| **2166** | **0.167** | **2250 (Late)** | **0.692** | **0.0, 0.004, 0.995 (Late)** |
| 2166 | 0.529 | 2250 (Late) | 0.810 | 0.002, 0.169, 0.829 (Late) |
| 2160 | 0.0 | 2250 (Late) | 1.0 | 0.008, 0.314, 0.679 (Late) |
| 2190 | 0.167 | 2250 (Late) | 0.818 | 0.002, 0.09, 0.908 (Late) |
| 2280 | 0.07 | 2460 (Late) | 0.286 | 0.001, 0.014, 0.986 (Late) |
| 2160 | 0.143 | 2250 (Late) | 1.0 | 0.002, 0.11, 0.888 (Late) |
| 2172 | 0.143 | 2250 (Late) | 0.8 | 0.175, 0.636, 0.189 (Mid) |
| 2190 | 0.143 | 2310 (Late) | 1.0 | 0.002, 0.111, 0.887 (Late) |
| 2190 | 0.0 | 2310 (Late) | 1.0 | 0.06, 0.673, 0.267 (Mid) |

**Table S5: Subgroup analysis.**
Subgroup analysis is conducted on Dataset A's test set. For each of the 11 clinical variables, the dataset is divided into 2 subgroups and then evaluated by FTXGB. N: the number of patients in the subgroup, ACC: macro-average accuracy, PR: macro-average precision, RE: macro-average recall, SP: macro-average specificity, F1: macro-average F1 score, AUROC: macro-average area under ROC curve, AUPR: macro-average area under precision-recall curve.

| Characteristics | N | ACC | PR | RE | SP | F1 | AUROC | AUPR |
|---|---|---|---|---|---|---|---|---|
| Age <=36 | 134 | 0.820 | 0.844 | 0.753 | 0.870 | 0.788 | 0.884 | 0.822 |
| Age >36 | 142 | 0.821 | 0.833 | 0.736 | 0.864 | 0.773 | 0.865 | 0.791 |
| Infertility duratioin <=4 | 220 | 0.809 | 0.813 | 0.741 | 0.860 | 0.771 | 0.889 | 0.815 |
| Infertility duratioin >4 | 54 | 0.852 | 0.796 | 0.819 | 0.896 | 0.805 | 0.895 | 0.850 |
| BMI <=22.3 | 139 | 0.813 | 0.804 | 0.741 | 0.861 | 0.768 | 0.900 | 0.833 |
| BMI >22.3 | 135 | 0.822 | 0.824 | 0.766 | 0.872 | 0.788 | 0.879 | 0.811 |
| Duration of stimulation<=10 | 121 | 0.777 | 0.729 | 0.670 | 0.821 | 0.700 | 0.839 | 0.744 |
| Duration of stimulation >10 | 153 | 0.850 | 0.865 | 0.801 | 0.895 | 0.827 | 0.920 | 0.869 |
| After trigger hCG <=80 | 138 | 0.848 | 0.862 | 0.776 | 0.883 | 0.810 | 0.889 | 0.823 |
| After trigger hCG >80 | 133 | 0.784 | 0.775 | 0.730 | 0.845 | 0.750 | 0.885 | 0.818 |
| hMG day <=8 | 125 | 0.784 | 0.768 | 0.669 | 0.820 | 0.708 | 0.836 | 0.753 |
| hMG day >8 | 149 | 0.846 | 0.854 | 0.803 | 0.896 | 0.824 | 0.920 | 0.869 |
| hMG dose per day is 225 | 148 | 0.824 | 0.813 | 0.737 | 0.861 | 0.769 | 0.874 | 0.798 |
| hMG dose per day not 225 | 126 | 0.810 | 0.805 | 0.788 | 0.869 | 0.794 | 0.900 | 0.838 |
| hMG dose <=2050 | 158 | 0.816 | 0.796 | 0.699 | 0.845 | 0.737 | 0.862 | 0.769 |
| hMG dose >2050 | 116 | 0.819 | 0.818 | 0.799 | 0.886 | 0.808 | 0.913 | 0.863 |
| MPA dose <=36 | 165 | 0.806 | 0.747 | 0.708 | 0.841 | 0.725 | 0.862 | 0.769 |
| MPA dose >36 | 119 | 0.834 | 0.871 | 0.806 | 0.896 | 0.827 | 0.911 | 0.863 |
| AFC <=5 | 129 | 0.814 | 0.706 | 0.858 | 0.862 | 0.753 | 0.907 | 0.915 |
| AFC >5 | 145 | 0.821 | 0.833 | 0.736 | 0.864 | 0.733 | 0.865 | 0.792 |
| Follicles <=9 | 176 | 0.807 | 0.704 | 0.681 | 0.847 | 0.689 | 0.873 | 0.742 |
| Follicles >9 | 98 | 0.837 | 0.834 | 0.868 | 0.885 | 0.847 | 0.887 | 0.880 |

**Table S6: Performance metrics for regression task of GnRH antagonist protocol, long protocol, short protocol and mild stimulation protocol.**
For each COH protocol, 10-fold cross-validation is performed on the corresponding dataset. FTLGB outperforms the other models across all 4 COH protocols. RMSE: Root Mean Squared Error, MAE: Mean Absolute Error, *r*: Pearson correlation coefficient.

|  | EN | RF | XGBoost | CatBoost | LightGBM | FTLGB |
|---|---|---|---|---|---|---|
| **GnRH antagonist protocol** | | | | | | |
| RMSE | 96.45 | 26.10 | 26.24 | 26.06 | 26.27 | 23.54 |
| MAE | 29.44 | 20.18 | 20.34 | 20.12 | 20.37 | 18.08 |
| *r* | 0.411 | 0.586 | 0.584 | 0.593 | 0.582 | 0.684 |
| **Long protocol** | | | | | | |
| RMSE | 52.17 | 40.93 | 41.28 | 41.53 | 40.93 | 36.67 |
| MAE | 42.44 | 32.83 | 33.02 | 33.27 | 32.79 | 29.31 |
| *r* | 0.507 | 0.731 | 0.726 | 0.724 | 0.732 | 0.792 |
| **Short protocol** | | | | | | |
| RMSE | 81.11 | 43.09 | 42.96 | 43.33 | 42.87 | 43.80 |
| MAE | 43.51 | 33.25 | 33.40 | 33.83 | 33.26 | 34.00 |
| *r* | 0.348 | 0.639 | 0.641 | 0.635 | 0.644 | 0.625 |
| **Mild stimulation protocol** | | | | | | |
| RMSE | 44.79 | 39.24 | 39.32 | 39.43 | 39.22 | 37.72 |
| MAE | 35.74 | 31.28 | 31.14 | 31.38 | 31.20 | 29.70 |
| *r* | 0.691 | 0.775 | 0.774 | 0.773 | 0.775 | 0.794 |

**Table S7: Total performance metrics for 3-class classification task of GnRH antagonist protocol, long protocol, short protocol and mild stimulation protocol.** For each COH protocol, 10-fold cross-validation is performed on the corresponding dataset. The 2 sub-tables show respectively the results of traditional machine learning methods and deep tabular models.

|  | SVM | RF | LightGBM | CatBoost | XGBoost |
|---|---|---|---|---|---|
| **GnRH antagonist protocol** | | | | | |
| ACC | 0.597 | 0.680 | 0.684 | 0.677 | 0.703 |
| PR | 0.594 | 0.663 | 0.671 | 0.667 | 0.697 |
| RE | 0.567 | 0.656 | 0.671 | 0.673 | 0.683 |
| SP | 0.782 | 0.835 | 0.839 | 0.836 | 0.846 |
| F1 | 0.569 | 0.657 | 0.668 | 0.666 | 0.683 |
| AUROC | 0.753 | 0.838 | 0.857 | 0.859 | 0.860 |
| AUPR | 0.613 | 0.714 | 0.750 | 0.751 | 0.751 |
| **Long protocol** | | | | | |
| ACC | 0.580 | 0.655 | 0.654 | 0.677 | 0.664 |
| PR | 0.585 | 0.659 | 0.655 | 0.667 | 0.662 |
| RE | 0.597 | 0.693 | 0.683 | 0.693 | 0.671 |
| SP | 0.786 | 0.830 | 0.827 | 0.832 | 0.823 |
| F1 | 0.586 | 0.650 | 0.657 | 0.663 | 0.660 |
| AUROC | 0.753 | 0.837 | 0.841 | 0.843 | 0.844 |
| AUPR | 0.580 | 0.708 | 0.723 | 0.727 | 0.726 |
| **Short protocol** | | | | | |
| ACC | 0.552 | 0.659 | 0.687 | 0.677 | 0.706 |
| PR | 0.538 | 0.625 | 0.650 | 0.640 | 0.665 |
| RE | 0.522 | 0.625 | 0.650 | 0.642 | 0.654 |
| SP | 0.768 | 0.828 | 0.842 | 0.838 | 0.847 |
| F1 | 0.503 | 0.620 | 0.646 | 0.637 | 0.655 |
| AUROC | 0.684 | 0.819 | 0.830 | 0.826 | 0.837 |
| AUPR | 0.509 | 0.664 | 0.675 | 0.675 | 0.684 |
| **Mild stimulation protocol** | | | | | |
| ACC | 0.632 | 0.684 | 0.690 | 0.689 | 0.689 |
| PR | 0.639 | 0.686 | 0.691 | 0.690 | 0.696 |
| RE | 0.639 | 0.696 | 0.703 | 0.702 | 0.697 |
| SP | 0.813 | 0.841 | 0.844 | 0.844 | 0.842 |
| F1 | 0.637 | 0.688 | 0.694 | 0.692 | 0.694 |
| AUROC | 0.811 | 0.857 | 0.863 | 0.861 | 0.864 |
| AUPR | 0.691 | 0.755 | 0.764 | 0.761 | 0.764 |

|  | MLP | STG | TabNet | SAINT | NODE | FT Transformer | FTXGB (Ours) |
|---|---|---|---|---|---|---|---|
| **GnRH antagonist protocol** | | | | | | | |
| ACC | 0.668 | 0.669 | 0.604 | 0.697 | 0.708 | 0.713 | 0.714 |
| PR | 0.660 | 0.655 | 0.593 | 0.687 | 0.699 | 0.705 | 0.703 |
| RE | 0.648 | 0.649 | 0.590 | 0.683 | 0.690 | 0.716 | 0.696 |
| SP | 0.828 | 0.828 | 0.799 | 0.844 | 0.849 | 0.857 | 0.852 |
| F1 | 0.648 | 0.650 | 0.587 | 0.682 | 0.691 | 0.703 | 0.697 |
| AUROC | 0.821 | 0.827 | 0.766 | 0.861 | 0.862 | 0.870 | 0.876 |
| AUPR | 0.688 | 0.699 | 0.613 | 0.747 | 0.755 | 0.758 | 0.769 |
| **Long protocol** | | | | | | | |
| ACC | 0.620 | 0.590 | 0.597 | 0.649 | 0.634 | 0.644 | 0.641 |
| PR | 0.623 | 0.593 | 0.591 | 0.653 | 0.637 | 0.645 | 0.644 |
| RE | 0.637 | 0.609 | 0.629 | 0.669 | 0.653 | 0.681 | 0.659 |
| SP | 0.807 | 0.791 | 0.798 | 0.822 | 0.813 | 0.824 | 0.816 |
| F1 | 0.627 | 0.597 | 0.600 | 0.660 | 0.655 | 0.640 | 0.648 |
| AUROC  | 0.805 | 0.789 | 0.783 | 0.831 | 0.830 | 0.833 | 0.834 |
| AUPR   | 0.653 | 0.634 | 0.613 | 0.709 | 0.701 | 0.711 | 0.713 |
| **Short protocol** | | | | | | | |
| ACC    | 0.627 | 0.652 | 0.622 | 0.680 | 0.679 | 0.663 | 0.688 |
| PR     | 0.587 | 0.609 | 0.586 | 0.644 | 0.638 | 0.632 | 0.649 |
| RE     | 0.584 | 0.612 | 0.589 | 0.639 | 0.632 | 0.633 | 0.639 |
| SP     | 0.808 | 0.822 | 0.810 | 0.837 | 0.833 | 0.834 | 0.840 |
| F1     | 0.581 | 0.607 | 0.581 | 0.636 | 0.631 | 0.625 | 0.638 |
| AUROC  | 0.770 | 0.796 | 0.767 | 0.819 | 0.818 | 0.821 | 0.826 |
| AUPR   | 0.602 | 0.630 | 0.592 | 0.661 | 0.656 | 0.665 | 0.669 |
| **Mild stimulation protocol** | | | | | | | |
| ACC    | 0.837 | 0.842 | 0.816 | 0.856 | 0.862 | 0.860 | 0.862 |
| PR     | 0.661 | 0.661 | 0.643 | 0.683 | 0.688 | 0.685 | 0.682 |
| RE     | 0.671 | 0.667 | 0.648 | 0.686 | 0.693 | 0.686 | 0.686 |
| SP     | 0.667 | 0.669 | 0.655 | 0.696 | 0.696 | 0.703 | 0.693 |
| F1     | 0.827 | 0.828 | 0.820 | 0.840 | 0.841 | 0.843 | 0.839 |
| AUROC  | 0.837 | 0.842 | 0.816 | 0.856 | 0.862 | 0.860 | 0.862 |
| AUPR   | 0.726 | 0.738 | 0.690 | 0.754 | 0.763 | 0.760 | 0.762 |

**Table S8: Student's t-test of optimal trigger-OPU interval between different COH protocols.**

The results indicate that the optimal trigger-OPU interval varied significantly among different COH protocols.

| Protocol 1 | Protocol 2 | Freedom degree | t-values |
|---|---|---|---|
| PPOS | GnRH antagonist | 9,914 | 25.52 ($P < 0.001$) |
| PPOS | short | 10,484 | 10.63 ($P < 0.001$) |
| PPOS | long | 10,968 | -20.74 ($P < 0.001$) |
| PPOS | Mild stimulation | 11,127 | 127.28 ($P = 0.0$) |
| GnRH antagonist | short | 3,634 | -10.41 ($P < 0.001$) |
| GnRH antagonist | long | 4,118 | -29.90 ($P < 0.001$) |
| GnRH antagonist | Mild stimulation | 4,277 | 60.76 ($P = 0.0$) |
| short | long | 4,688 | -19.28 ($P < 0.001$) |
| short | Mild stimulation | 4,847 | 69.53 ($P = 0.0$) |
| long | Mild stimulation | 5,331 | 91.40 ($P = 0.0$) |

**Table S9: Performance metrics of premature ovulation risk predictor on Dataset B's test set.**

The premature ovulation risk predictor is based on a XGBoost binary classifier. ACC: macro-average accuracy, PR: macro-average precision, RE: macro-average recall, SP: macro-average specificity, F1: macro-average F1 score, AUROC: macro-average area under ROC curve, AUPR: macro-average area under precision-recall curve.

| ACC | PR | RE | F1 | SP | AUROC | AUPR |
|---|---|---|---|---|---|---|
| 0.825 | 0.964 | 0.675 | 0.794 | 0.975 | 0.838 | 0.884 |

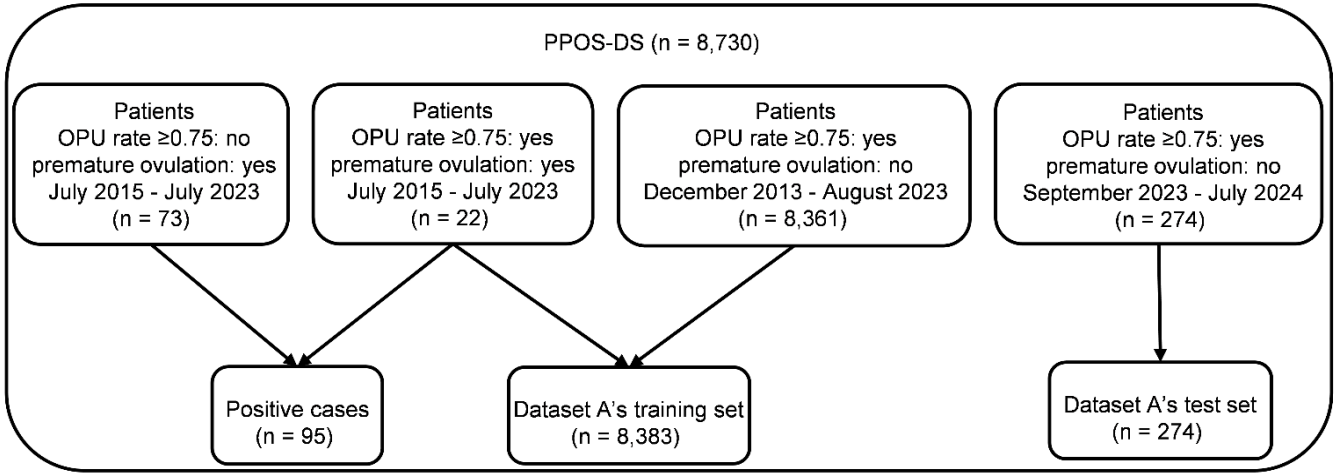

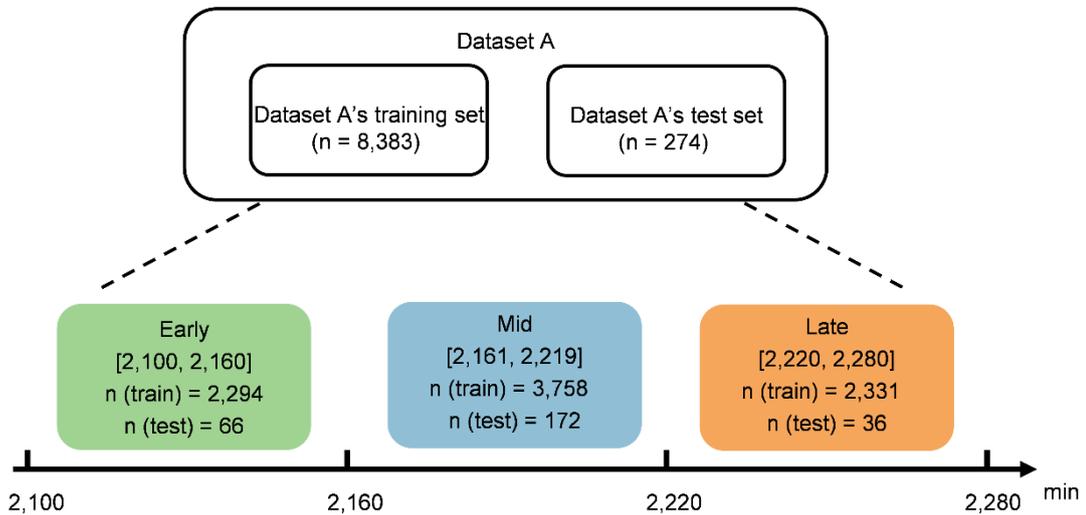

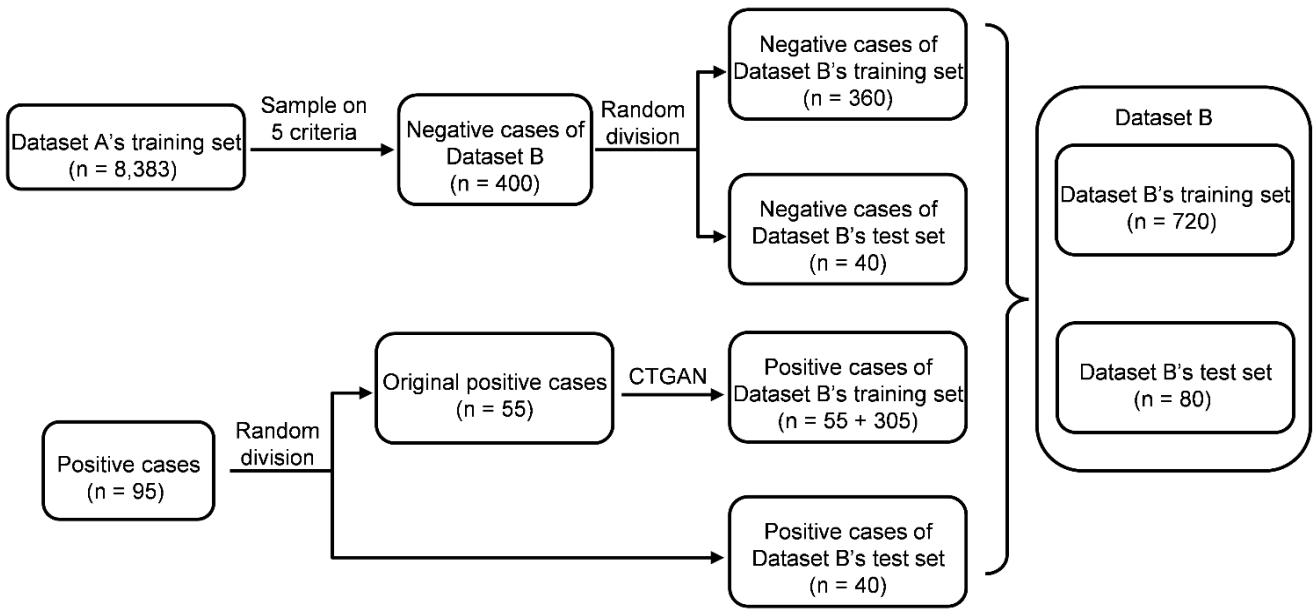

**Fig. S1: Composition of PPOS-DS and construction of datasets used for model training.**

**a,** Composition of PPOS-DS. Patients in PPOS-DS are categorized into four groups based on OPU rate (≥0.75), premature ovulation and outpatient visit date. These groups are further combined as shown to generate positive cases for Dataset B, Dataset A's training set and Dataset A's test set. **b,** Construction of Dataset A. Dataset A is used for training the optimal trigger-OPU interval predictors, comprising a training set and a test set. Each case has a label for the regression task (the original interval) and another for the 3-class classification task (the interval rank). Specifically, cases are relabeled by linearly splitting the original intervals into 3 ranks: Early (2,100-2,160 min), Mid (2,161-2,219 min) and Late (2,220-2,280 min). **c,** Construction of Dataset B. CTGAN refers to generating 305 new positive cases based on 55 original positive cases using CTGAN. 5 criteria: (1) not labeled as premature ovulation, (2) OPU rate =1, (3) 37.8 h≥ trigger-OPU interval ≥36.4 h, (4) antral follicle count >5, (5) E2 ratio>1.

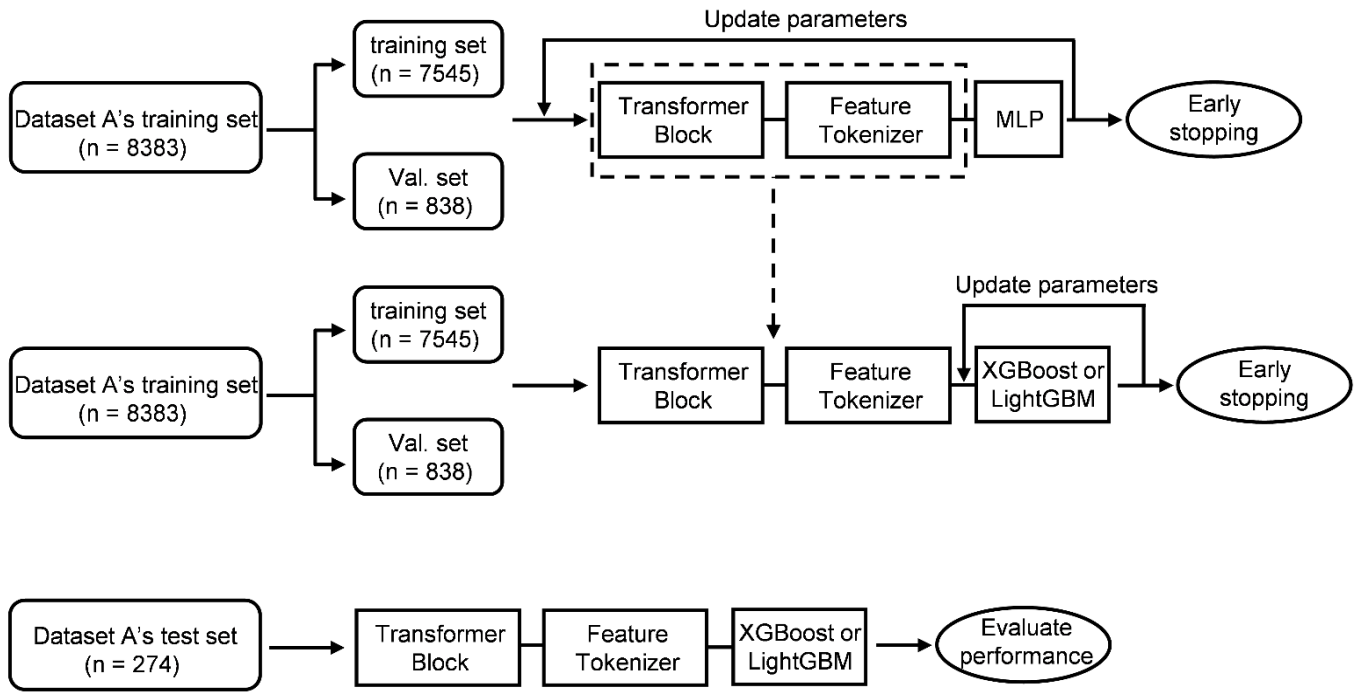

**Fig. S2: Construction of FTXGB and FTLGB.**
In the process of model training, Dataset A's training set is split into a training set and validation set with a 9:1 ratio. The validation set is used for Early stopping. To construct FTXGB, a normal FT Transformer (Feature Tokenizer + Transformer Block + MLP) is trained as a 3-class classification task initially. Weighted Cross-Entropy is used as the loss function in this step. Next, FTXGB is trained by retaining the trained Feature Tokenizer and Transformer Block in the normal FT Transformer and substituting MLP with XGBoost. Weighted Cross-Entropy is used here as well. FTLGB is constructed in a similar way that retain Feature Tokenizer and Transformer Block in the normal FT Transformer and substitute MLP with LightGBM to be trained. Negative Mean Squared Error is used as the loss function for the regressor. Finally, FTXGB and FTLGB are evaluated on Dataset A's test set.

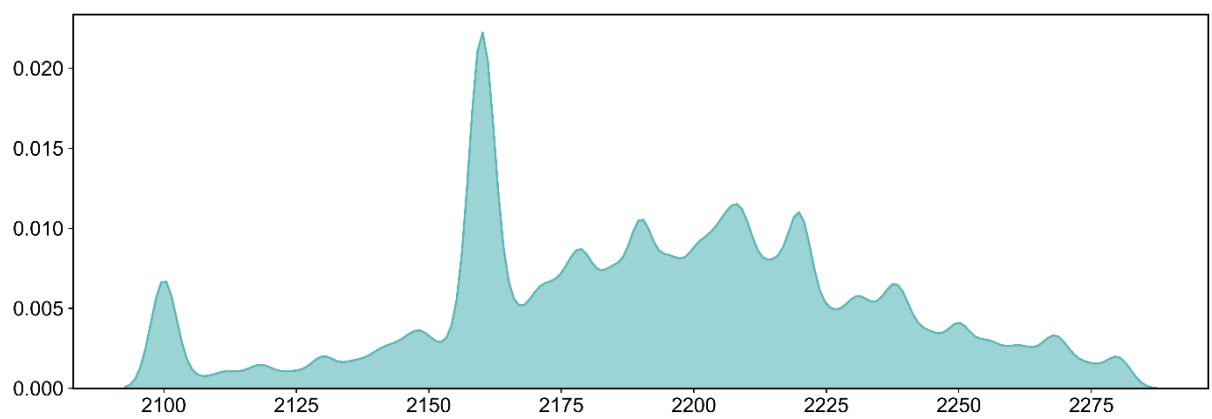

**Fig. S3: Distribution of label values in Dataset A's training set.**

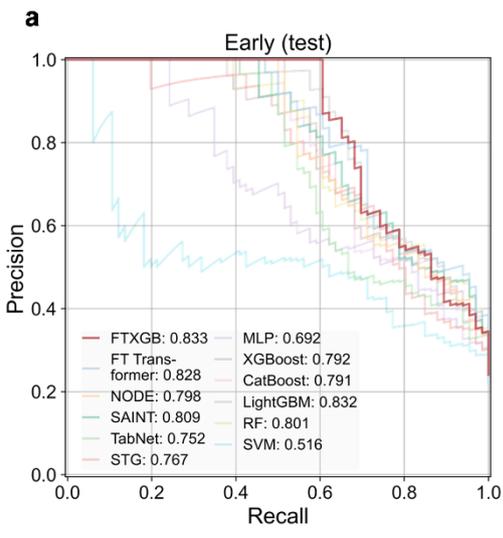 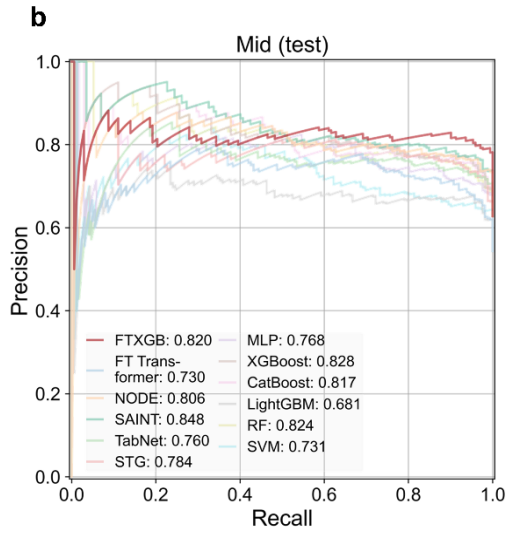 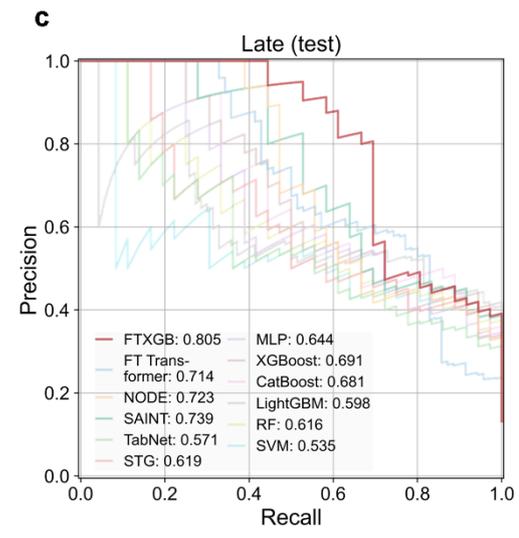
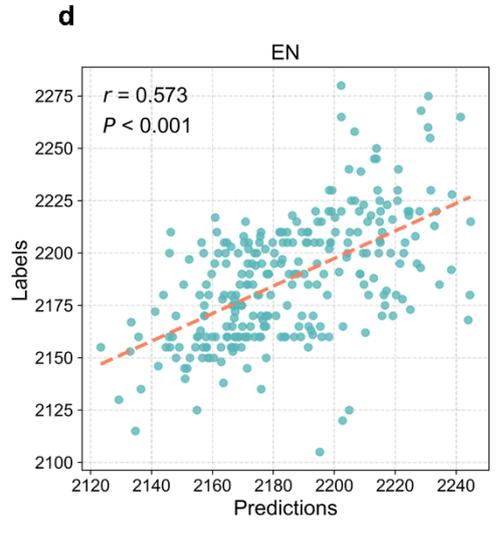 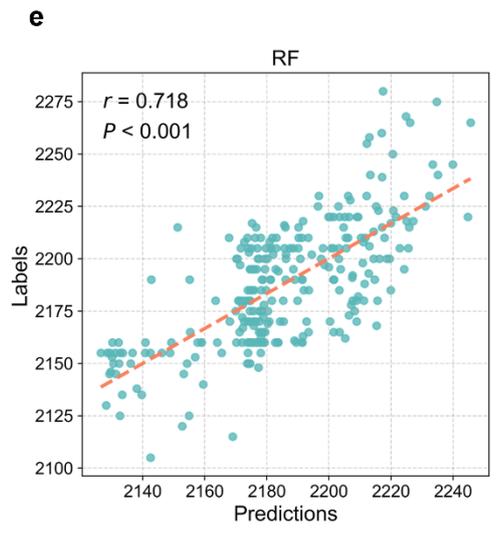 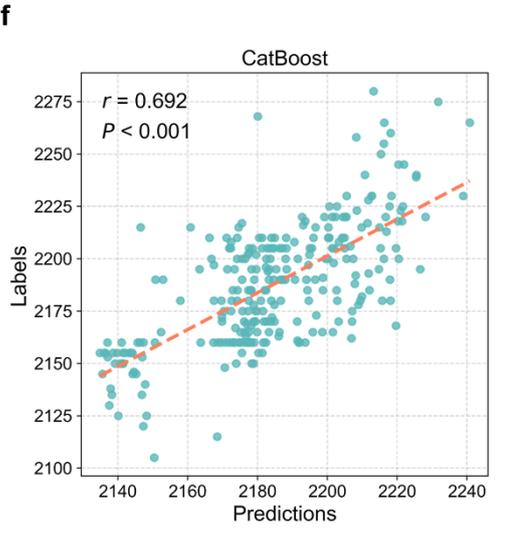
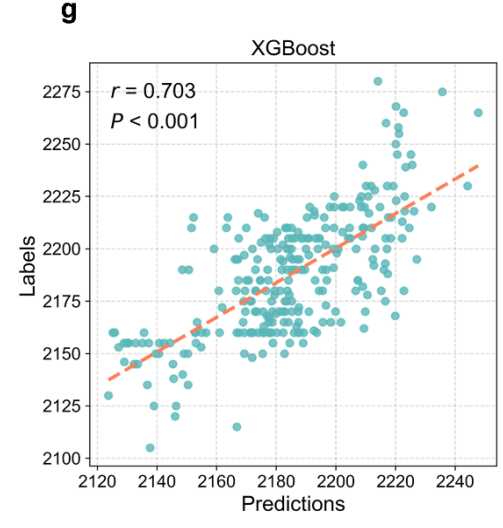 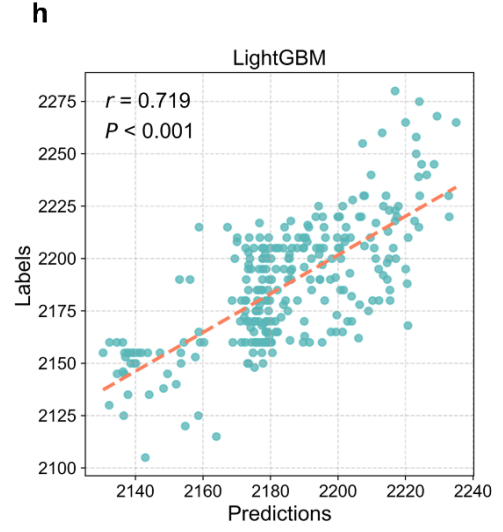 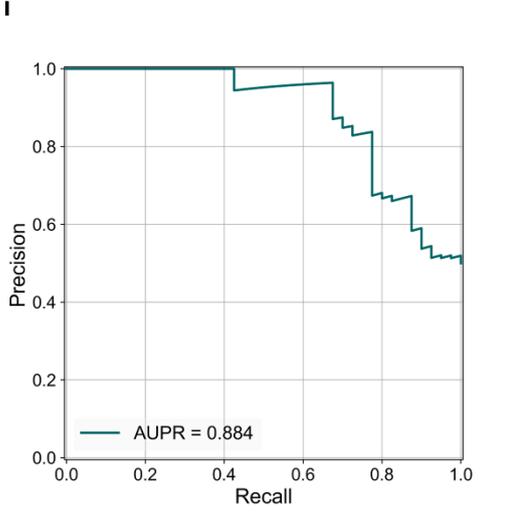
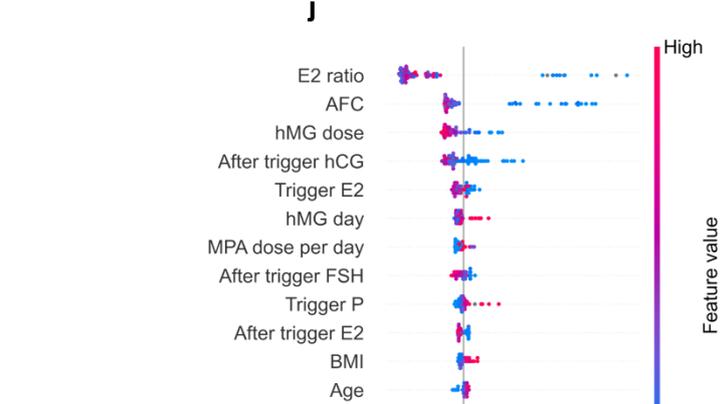 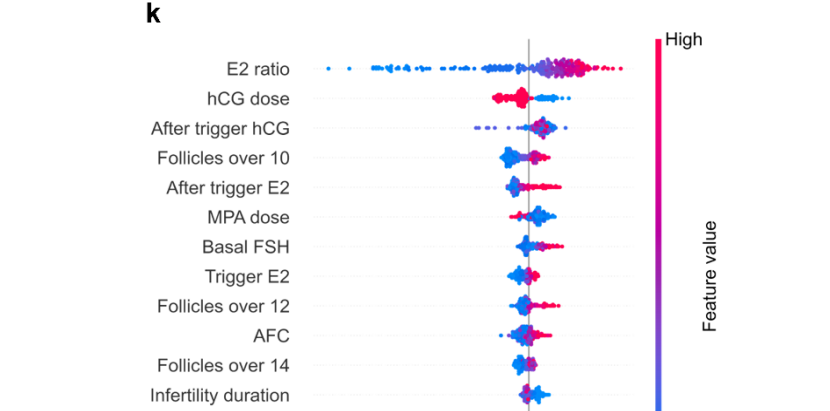

**Fig. S4: Model performance and interpretability.**
**a-c**, Precision-Recall Curve for each rank on Dataset A's test set across 12 models. **d-h**, Performance of 5 traditional machine learning models on Dataset A's test set. *r*, Pearson correlation coefficient. **i**, Precision-Recall Curve of the premature ovulation risk predictor on Dataset B's test set. **j**, Model interpretability of FTLGB by SHAP. **k**, model interpretability of the premature ovulation risk predictor (XGBoost) by SHAP. Top-15 most important clinical variables are shown in the left column. Red dots represent higher feature values and blue dots represent lower feature values. A positive SHAP value indicates a higher prediction value (**j**) or an increased probability of premature ovulation (**k**), and vice versa. The figure serves as a supplement to Fig. 2.

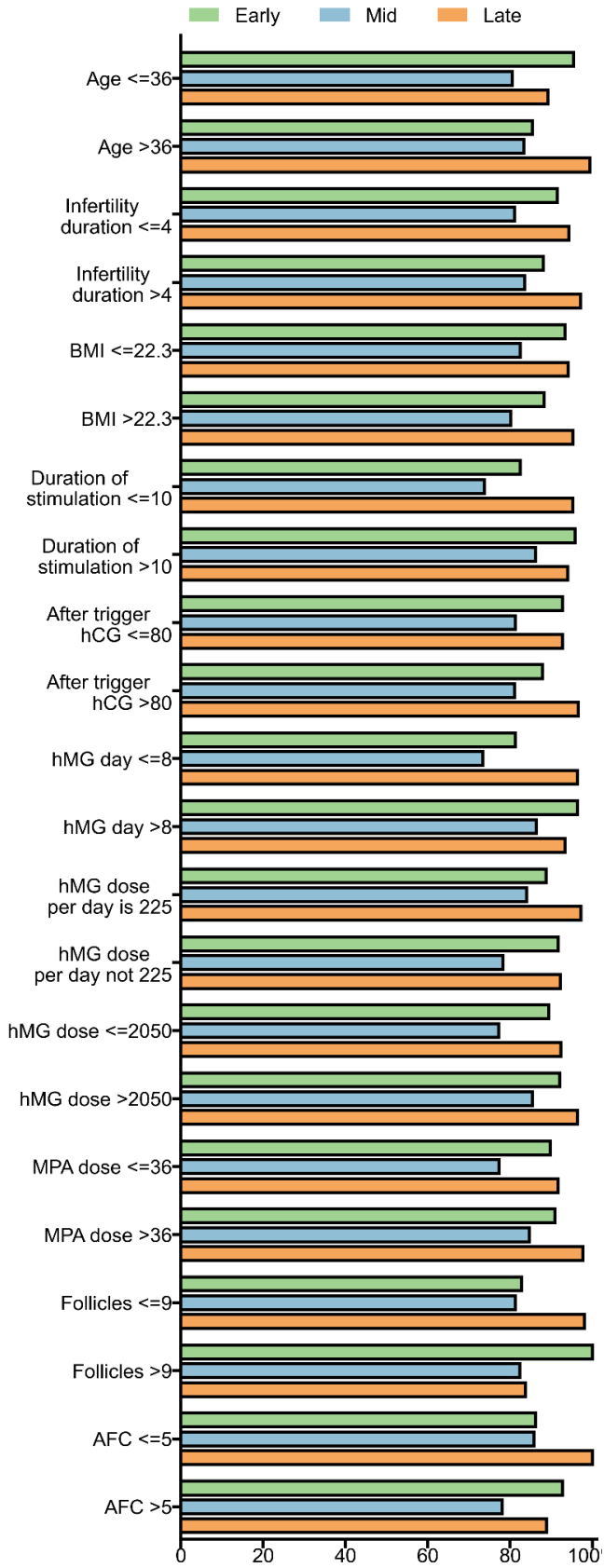
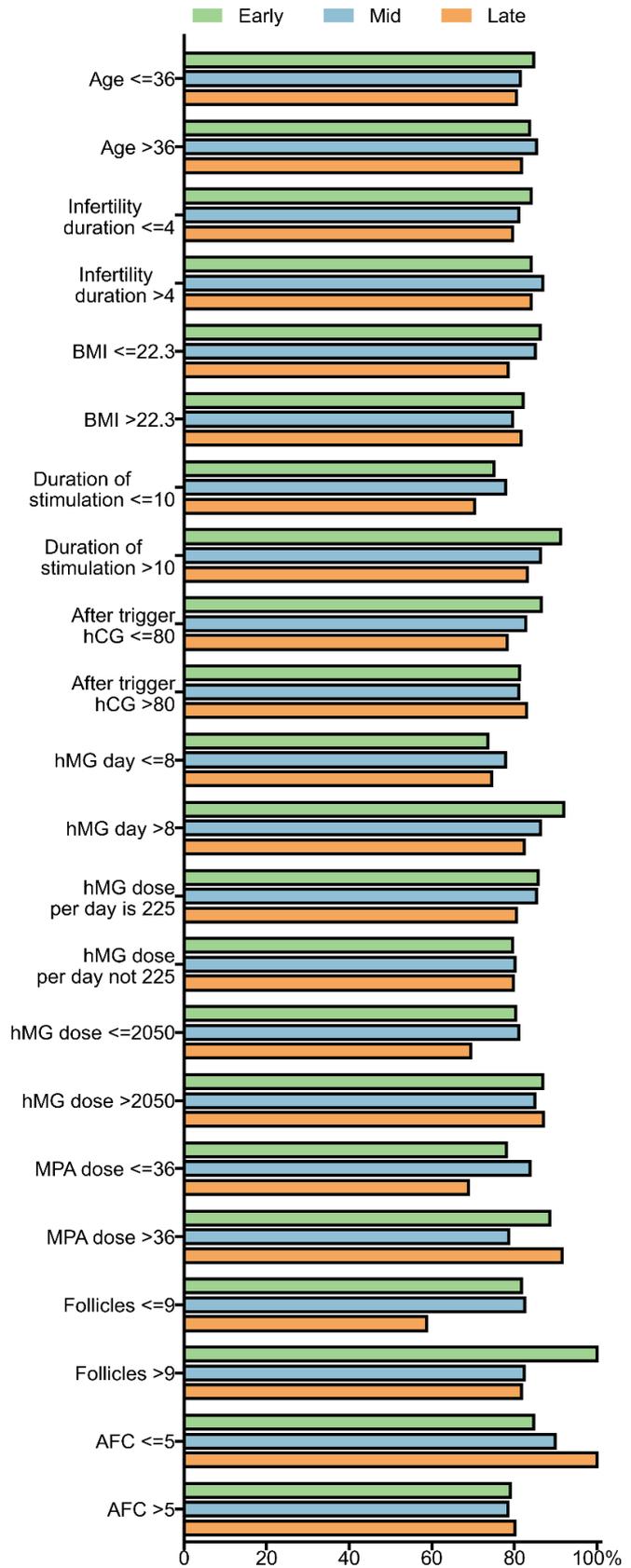

**Fig. S5: Complete results of subgroup analysis across 3 ranks.**

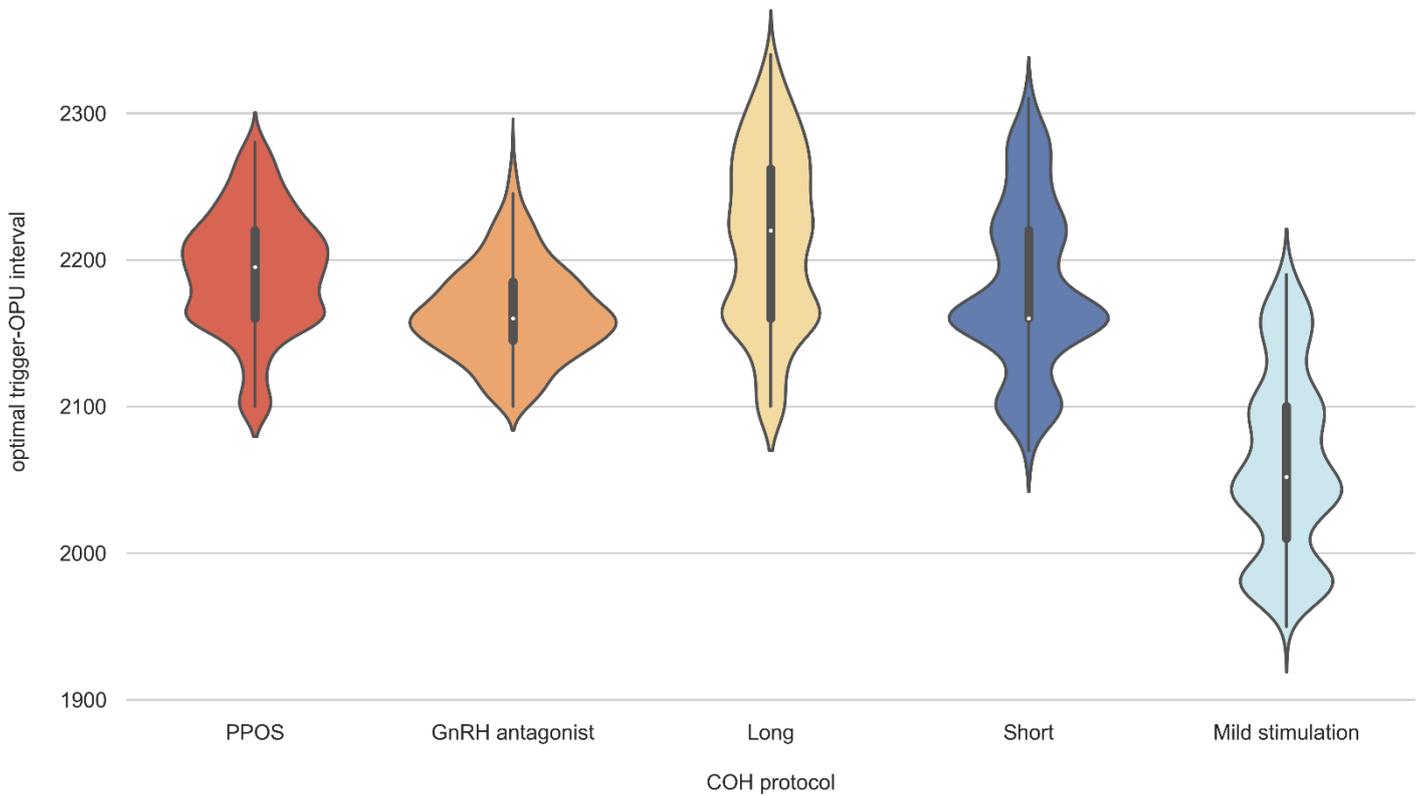

**Fig. S6: Distribution of the optimal trigger-OPU intervals across 5 COH protocols.**
Each violin plot is with a center line. Box limits and whiskers represent the median, interquartile range and minima/maxima within each COH protocol, respectively. PPOS protocol (n = 8,383), GnRH antagonist protocol (n = 1,533), long protocol (n = 2,587), short protocol (n = 2,103) and mild stimulation protocol (n = 2,746)

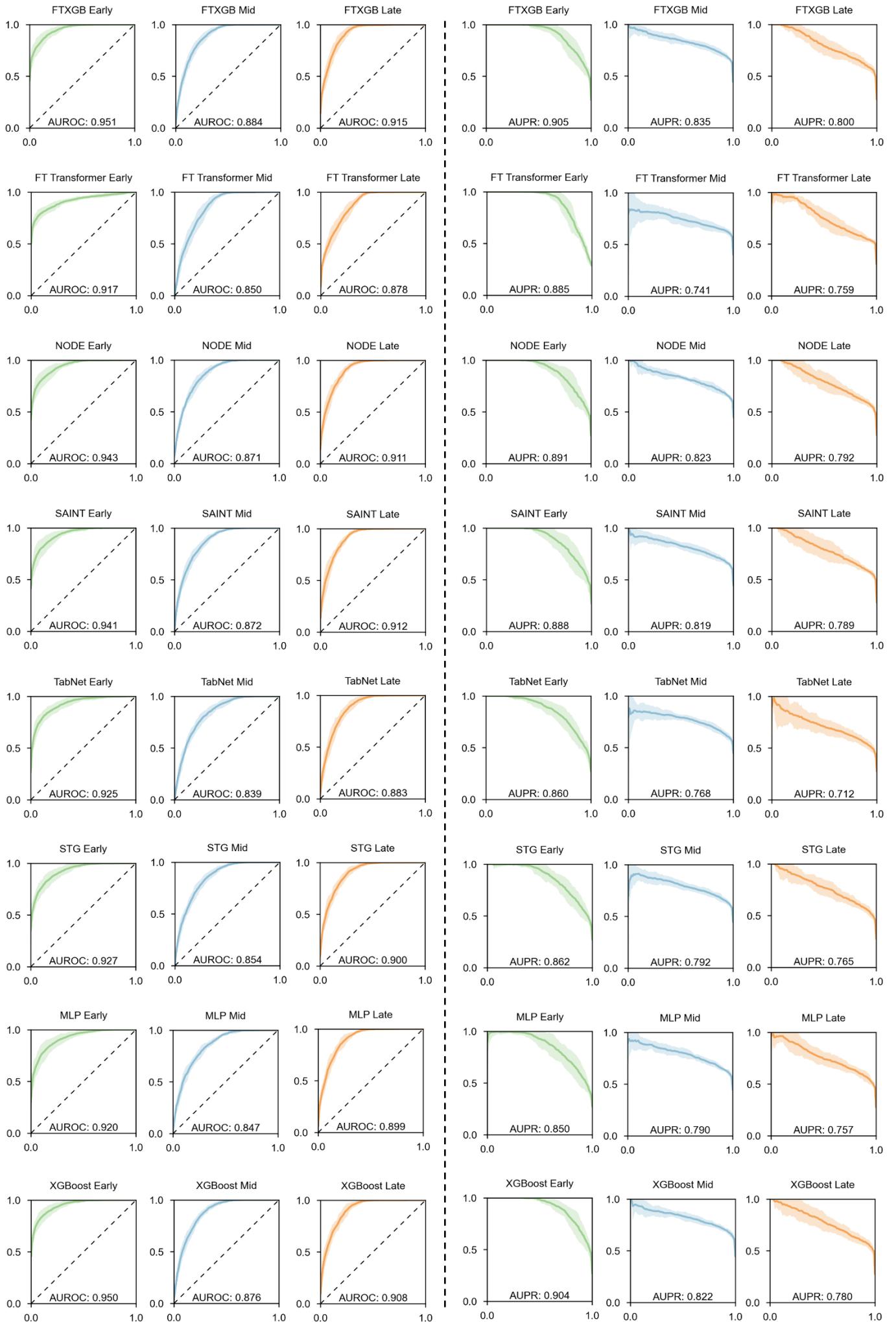

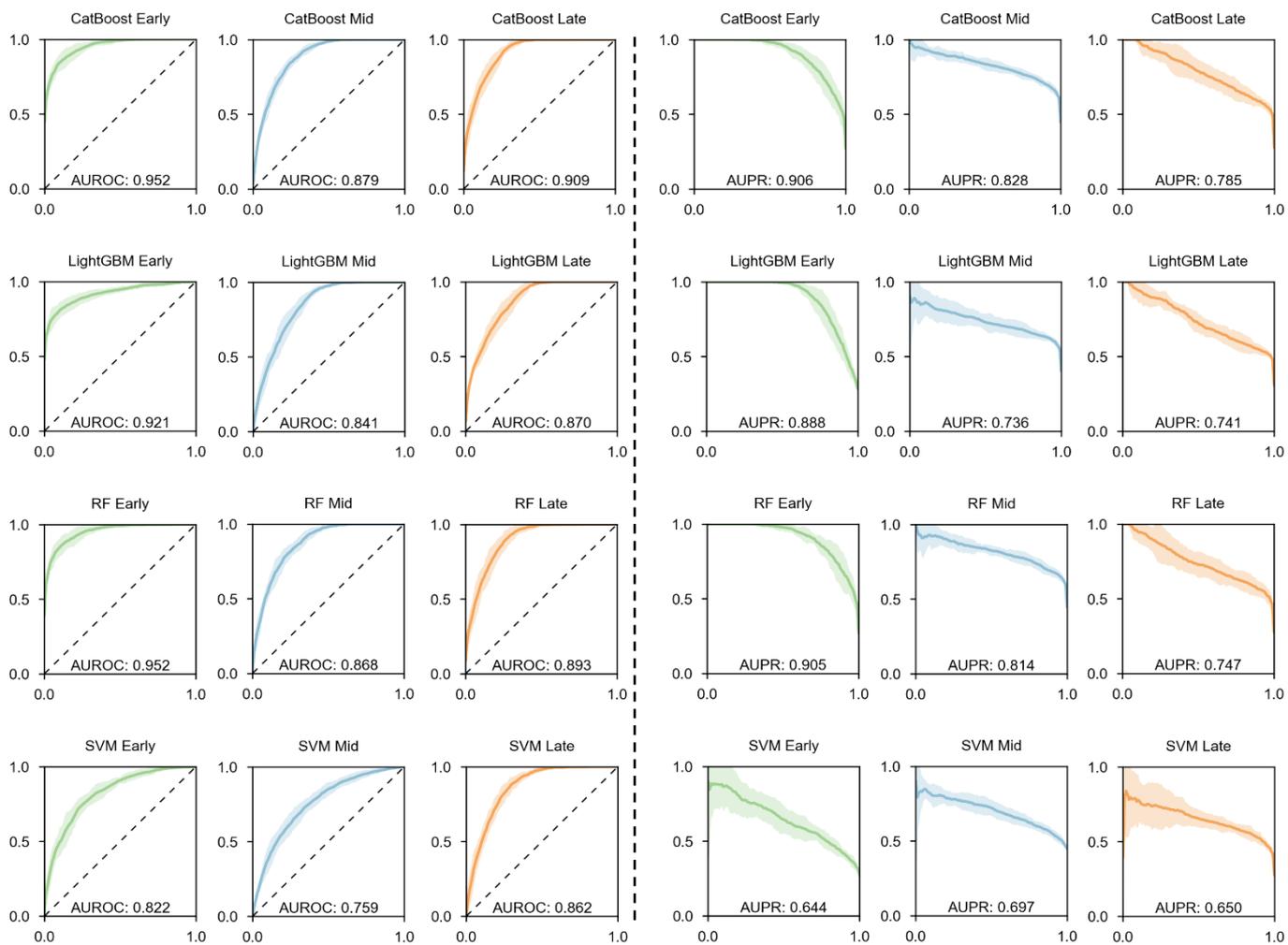

**Fig S7: ROC and Precision-Recall Curve of 12 classifiers across 3 ranks in 10-fold cross validation for PPOS protocol.**
The error bars show the 95% CI of the mean estimate.

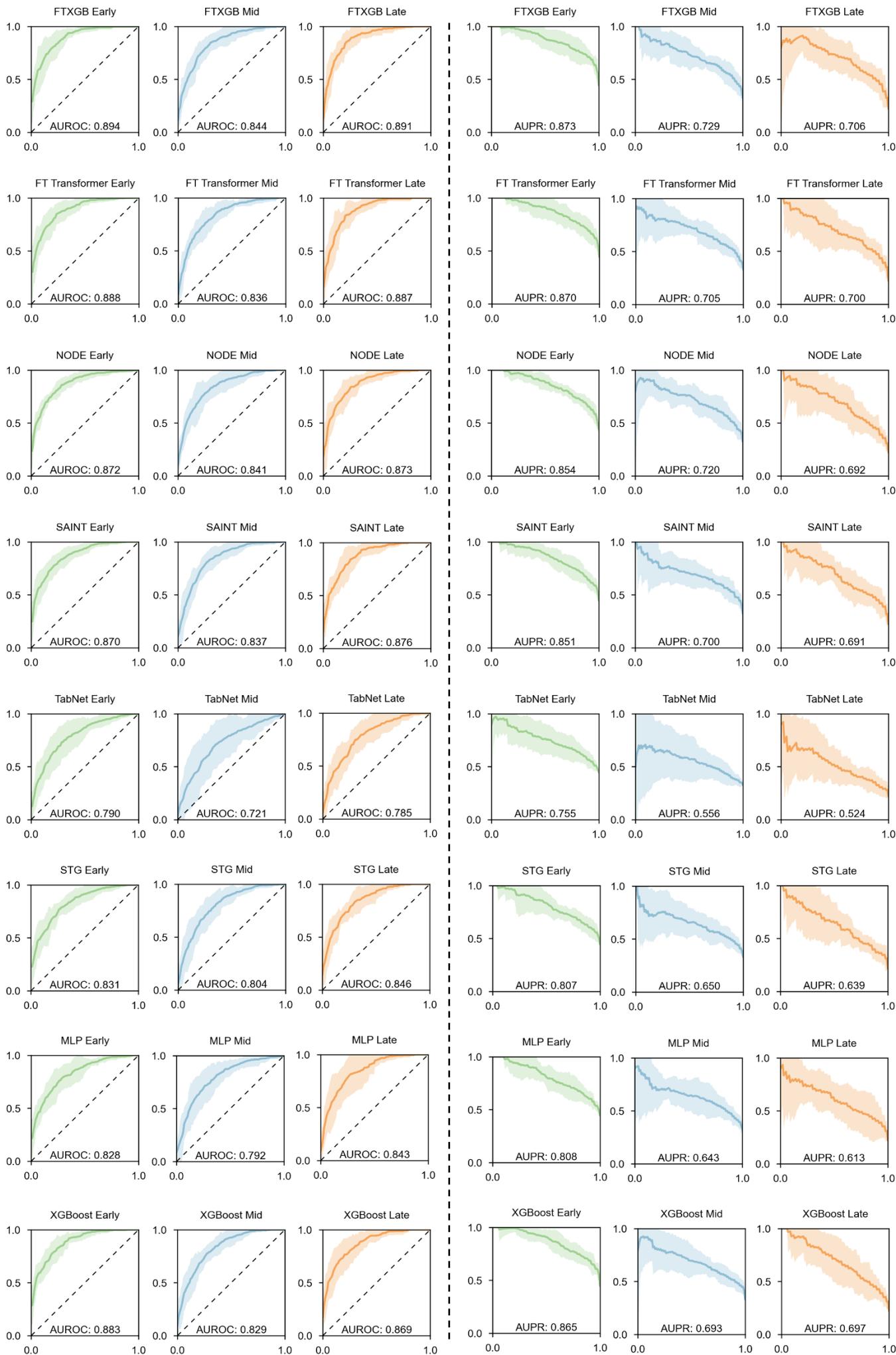

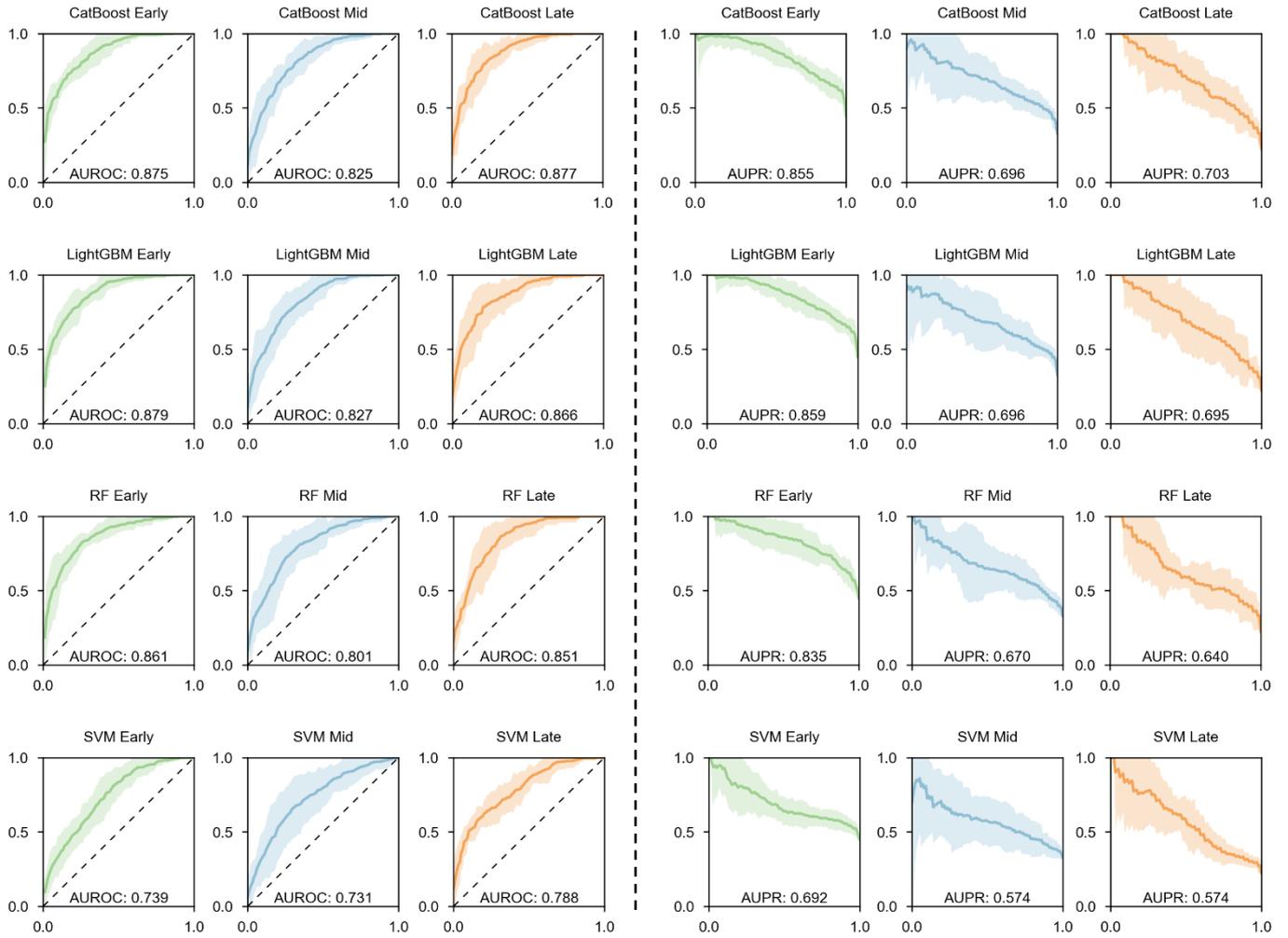

**Fig S8: ROC and Precision-Recall Curve of 12 classifiers across 3 ranks in 10-fold cross validation for the GnRH antagonist protocol.**
The error bars show the 95% CI of the mean estimate.

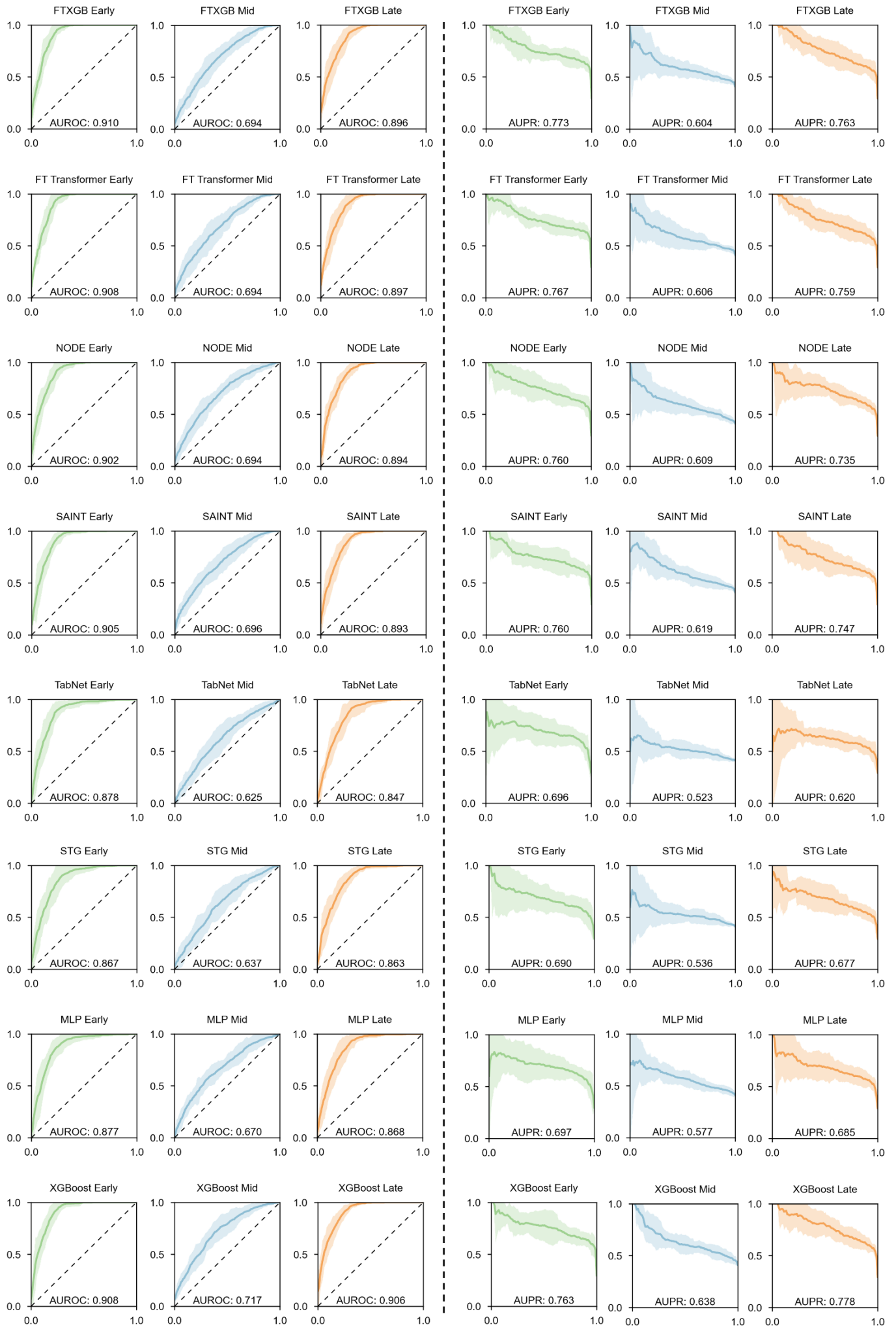

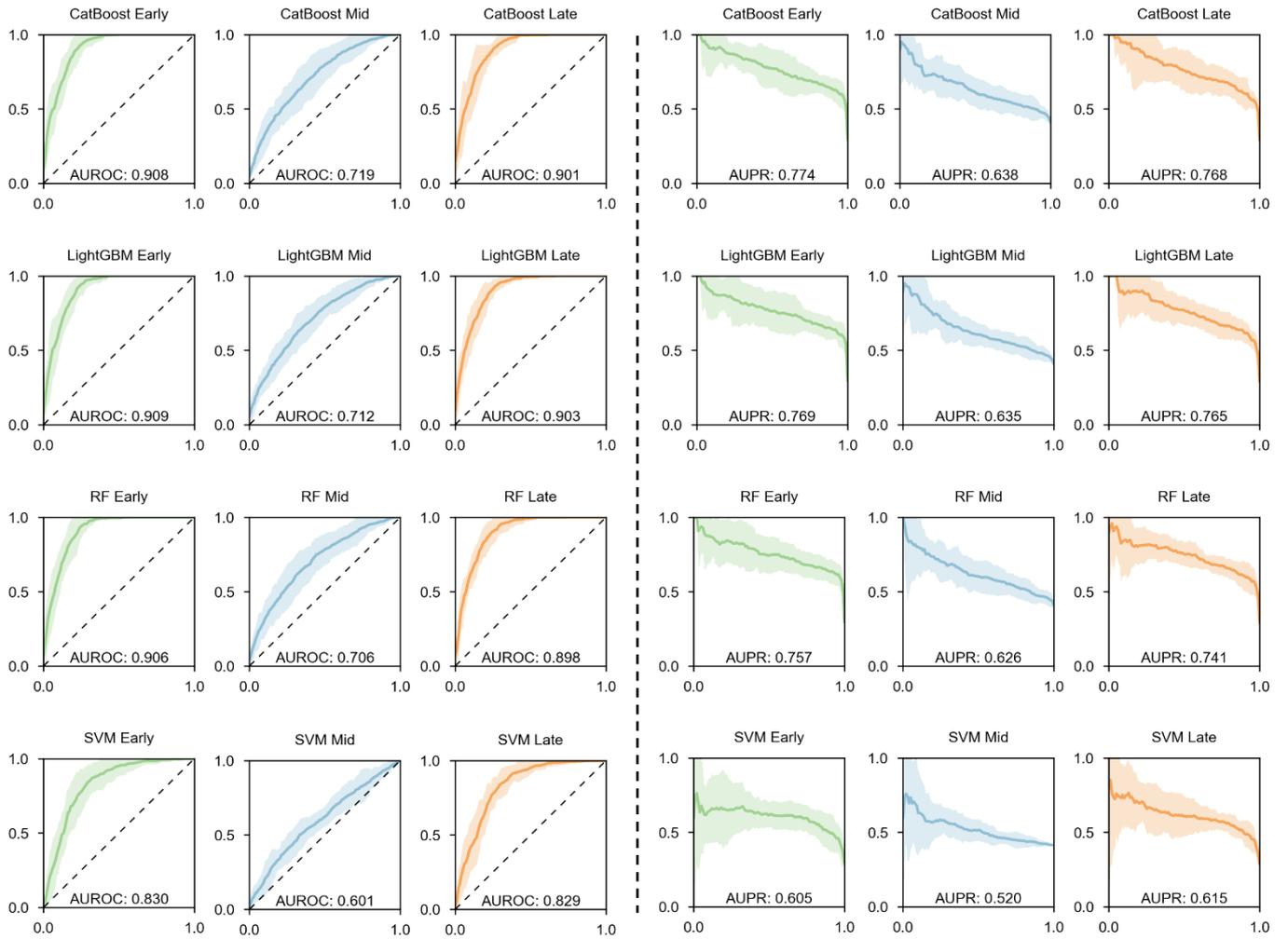

**Fig S9: ROC and Precision-Recall Curve of 12 classifiers across 3 ranks in 10-fold cross validation for the long protocol.**
The error bars show the 95% CI of the mean estimate.

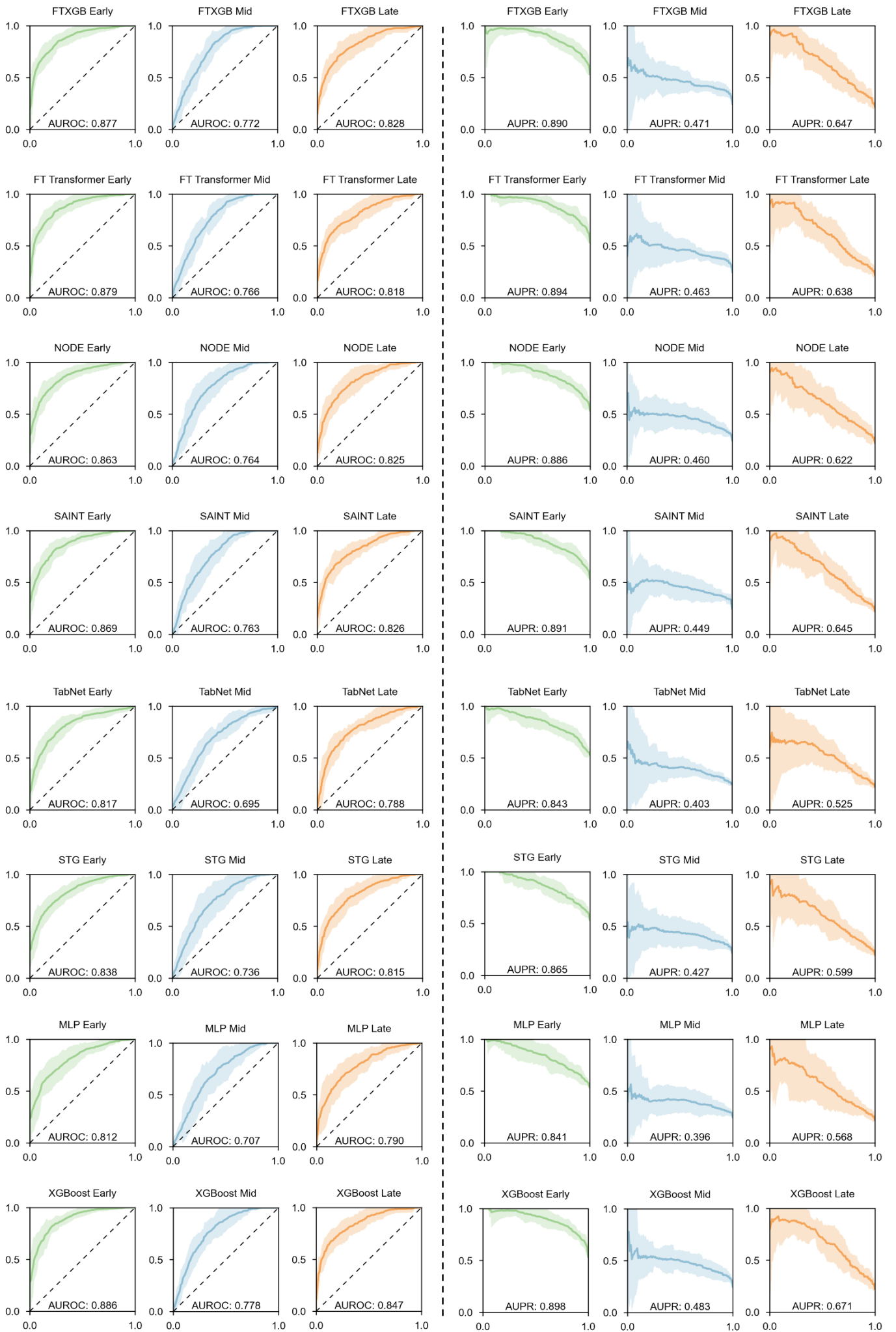

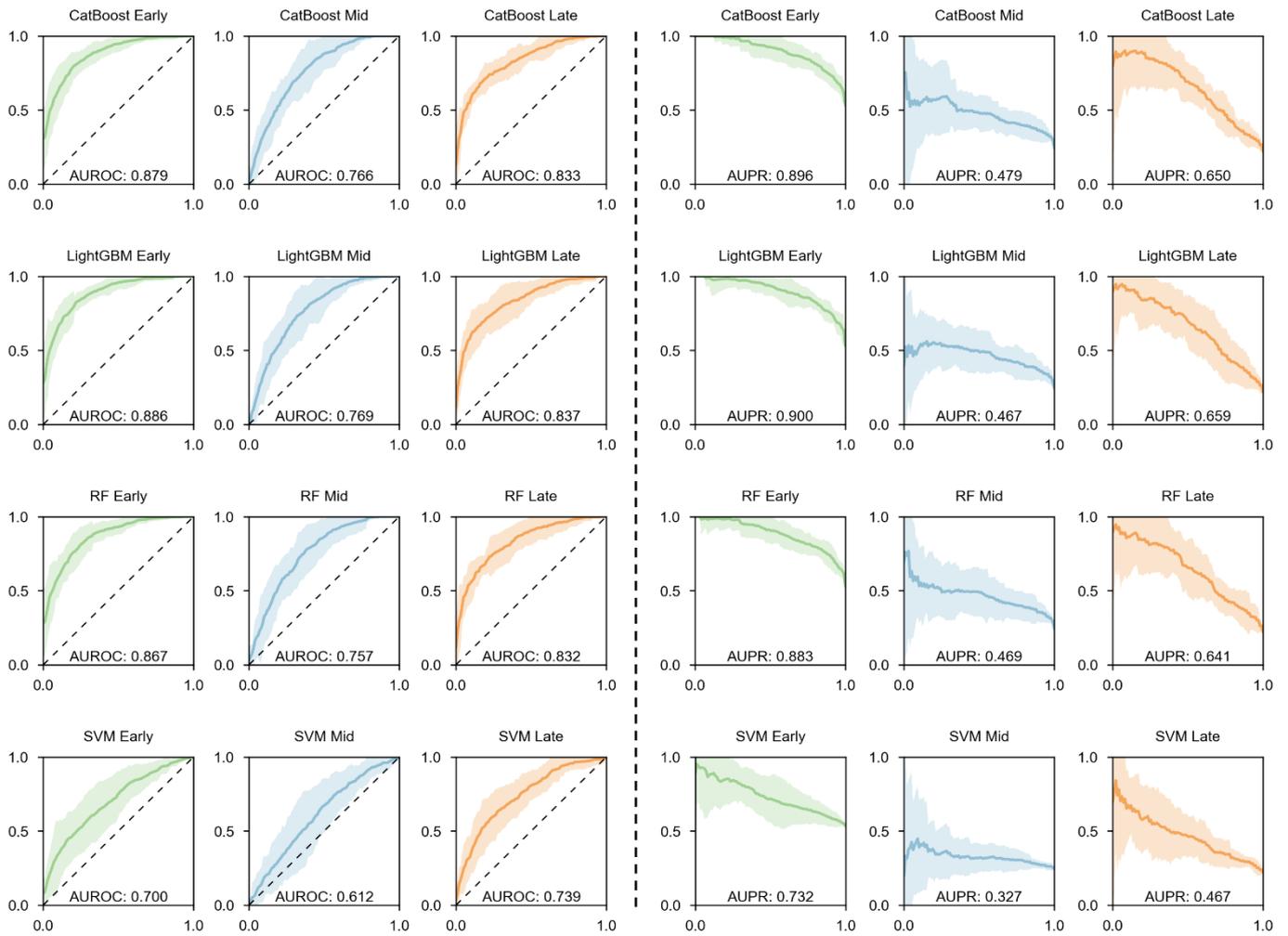

**Fig S10: ROC and Precision-Recall Curve of 12 classifiers across 3 ranks in 10-fold cross validation for the short protocol.**
The error bars show the 95% CI of the mean estimate.

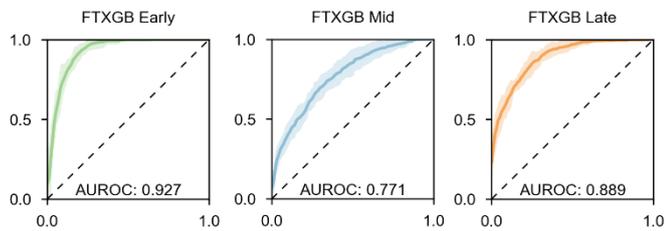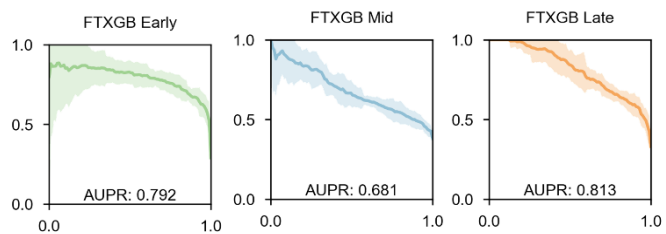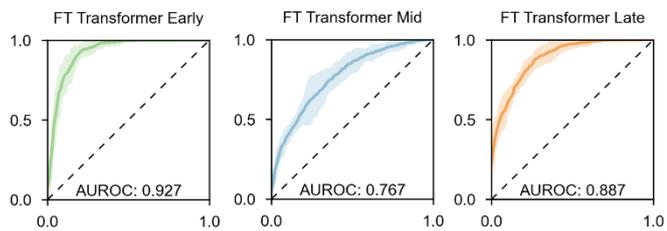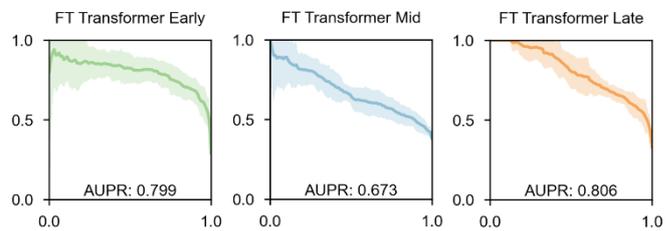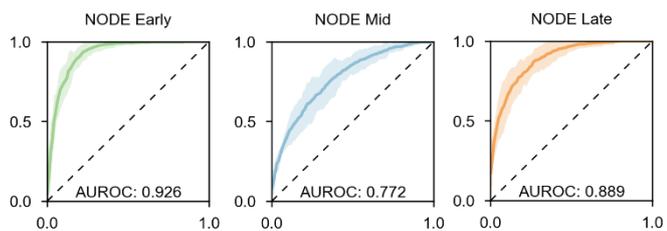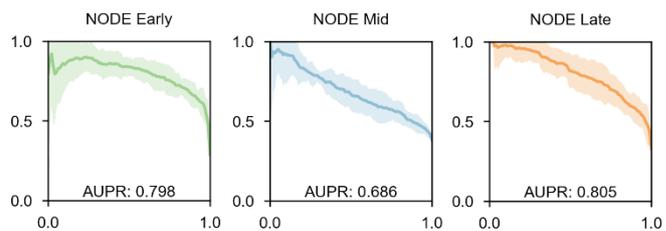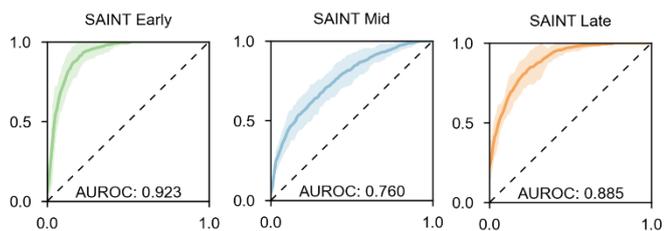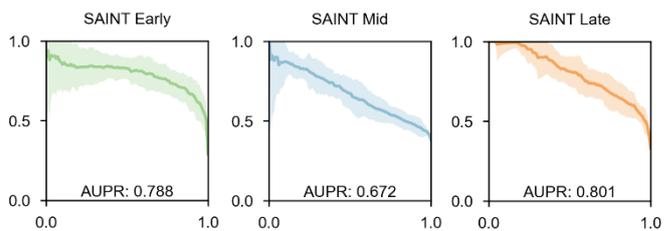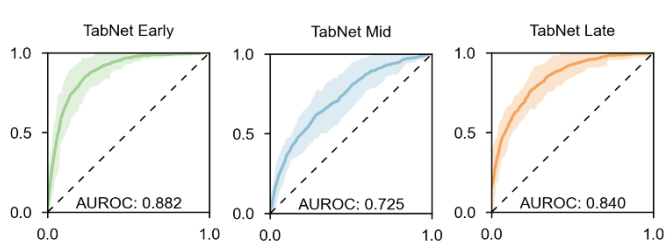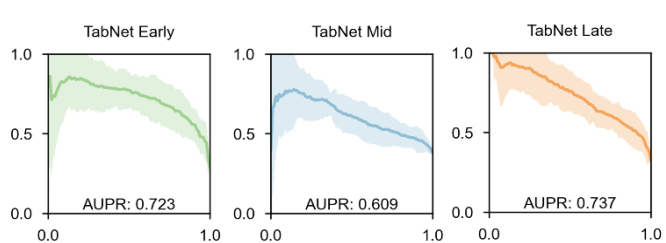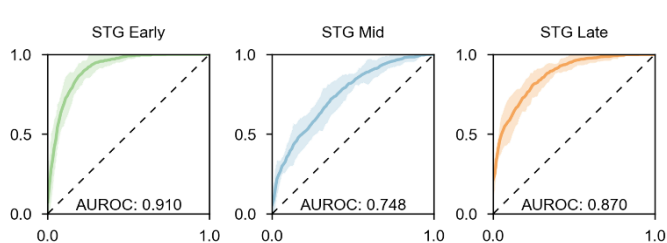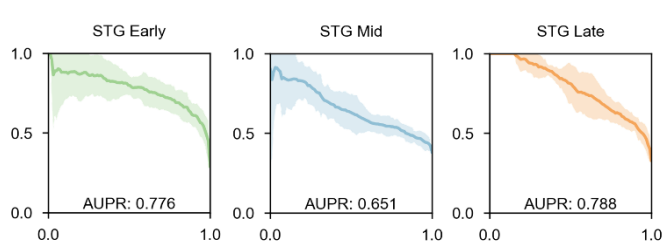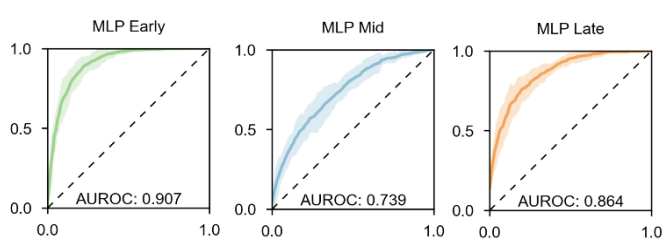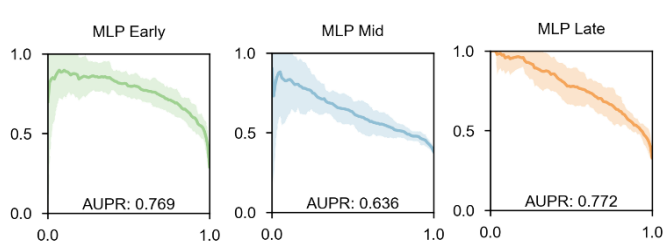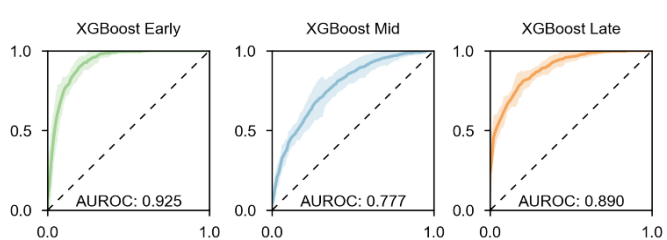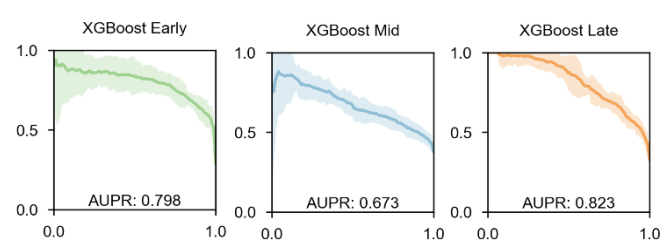

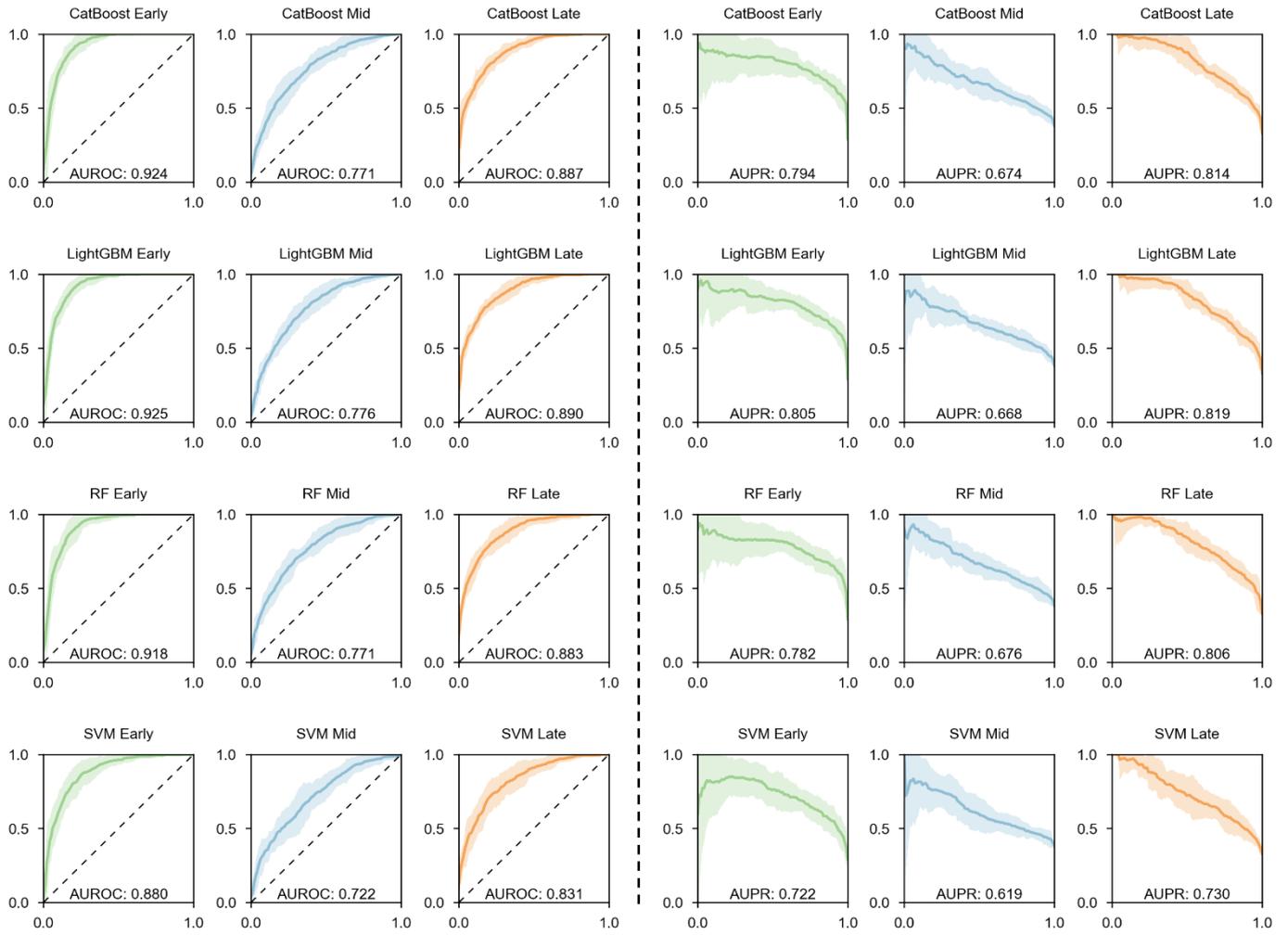

**Fig S11: ROC and Precision-Recall Curve of 12 classifiers across 3 ranks in 10-fold cross validation for the mild stimulation protocol.**
The error bars show the 95% CI of the mean estimate.

## Job Submitted Parameters

Age: 30 *Year*  
Duration of infertility: No info.  
BMI: 33.87 *kg/m²*

Duration of stimulation: 11 *day*  
Basal FSH: 3.8 *mIU/mL*  
Basal LH: 2.41 *mIU/mL*

Basal E2: 55 *pg/mL*  
Basal P: 0.1 *ng/mL*  
LH on the day of follow-up: 1.52 *mIU/mL*

E2 on the day of follow-up: 4501 *pg/mL*  
FSH on day of trigger: 13.48 *mIU/mL*  
LH on day of trigger: 1.52 *mIU/mL*

E2 on day of trigger: 4501 *pg/mL*  
P on day of trigger: 0.4 *ng/mL*  
hCG on the day after trigger: 14.85 *mIU/mL*

FSH on the day after trigger: 14.46 *mIU/mL*  
LH on the day after trigger: 15.99 *mIU/mL*  
E2 on the day after trigger: 5000 *pg/mL*

P on the day after trigger: 3.5 *ng/mL*  
Duration of HMG stimulation: 9 *Day*  
Total hMG dose: 3150 *IU*

Total hCG dose: 2000 *IU*  
Total MPA dose: 36 *mg*  
Antral follicles counts: 10

No. of follicles (diameter≥10mm): 15  
No. of follicles (diameter≥12mm): 15  
No. of follicles (diameter≥14mm): 13

No. of follicles (diameter≥16mm): 12  
No. of follicles (diameter≥18mm): 6  
No. of follicles (diameter≥20mm): 2

## Prediction Result

| Interval Value | Early Rate | Mid Rate | Late Rate |
|---|---|---|---|
| 37Hour 12Min | 0 | 0.005 | 0.995 |

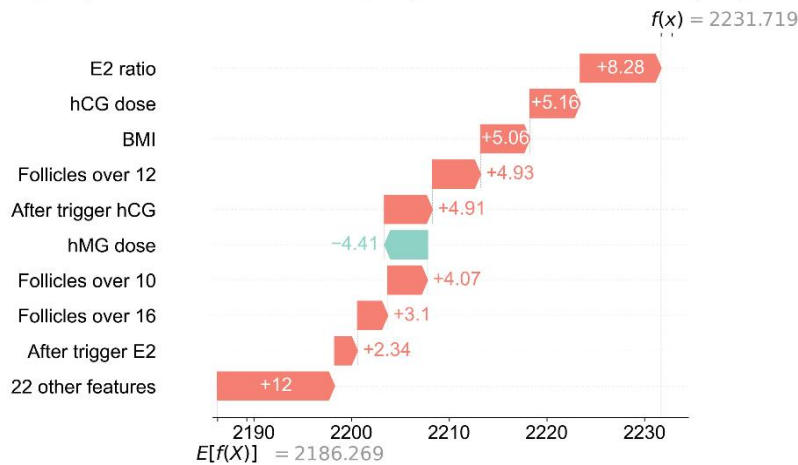

**Fig S12: Original ILETIA web server output.**
Original output of the specific case analyzed in Results